\documentclass[12pt,preprint]{aastex}

\slugcomment{submitted to ApJ}

\shorttitle{The lithium abundances for a large sample of red giants}
\shortauthors{Liu et al.}

\begin{document}


\title{The Lithium abundances of a large sample of red giants}


\author{
Y.J. Liu,\altaffilmark{1},  K. F. Tan\altaffilmark{1}, L. Wang
\altaffilmark{1}, G. Zhao,\altaffilmark{1}, Bun'ei
Sato,\altaffilmark{2},    Y. Takeda\altaffilmark{3}, H.N. Li,
\altaffilmark{1}  }


 \altaffiltext{1}{Key Laboratory of Optical Astronomy, National Astronomical
Observatories, Chinese Academy of Sciences, A20 Datun Road, Chaoyang
District, Beijing 100012; lyj@nao.cas.cn, gzhao@nao.cas.cn}

 \altaffiltext{2}{Tokyo Institute of Technology, 2-12-1 Ookayama, Meguro-ku,
Tokyo 152-8551, Japan}

 \altaffiltext{3}{National Astronomical Observatory of Japan, 2-21-1
 Osawa, Mitaka, Tokyo 181-8588, Japan}


\begin{abstract}
The lithium abundances for 378 G/K giants are derived with non-LTE
correction considered. Among these, there are 23 stars that host
planetary systems. The lithium abundance is investigated, as a
function of metallicity, effective temperature, and rotational
velocity, as well as the impact of a giant planet on G/K giants. The
results show that the lithium abundance is a function of metallicity
and effective temperature. The lithium abundance has no correlation
with rotational velocity at vsini $<$ 10 km s$^{-1}$. Giants with
planets present lower lithium abundance and slow rotational velocity
(vsini $<$ 4 km s$^{-1}$). Our sample includes three Li-rich G/K
giants, 36 Li-normal stars and 339 Li-depleted stars. The fraction
of Li-rich stars in this sample agrees with the general rate of less
than 1$\%$ in literature, and the stars that show normal amounts of
Li are supposed to possess the same abundance at the current
interstellar medium. For the Li-depleted giants, Li deficiency may
have already taken place at the main sequence stage for many
intermediate-mass (1.5-5 M$_{\odot}$) G/K giants. Finally, we
present the lithium abundance and kinematic parameters for an
enlarged sample of 565 giants using a compilation of literature, and
confirm that the lithium abundance is a function of metallicity and
effective temperature. With the enlarged sample, we investigate the
differences between the lithium abundance in thin-/thick-disk
giants, which indicate that the lithium abundance in thick-disk
giants is more depleted than that in thin-disk giants.

\end{abstract}


\section{Introduction}
Lithium is an important element in the understanding of chemical
evolution history of the Galaxy, as well as in the study of the
mixing process in stellar interiors. As is known, the lithium
abundance is believed to be a function of metallicity, effective
temperature, stellar mass, age, stellar rotation and chromospheric
activity. The behavior of lithium as a function of effective
temperature has been well studied both for subgiants (De Medeiros et
al. 1997; L$\grave{e}$bre et al. 1999; Randich et al. 2000; Mallik
et al. 2003) and giants (Brown et al. 1989; Wallerstein et al. 1994;
De Medeiros et al. 2000; de Laverny et al. 2003, L$\grave{e}$bre et
al. 2006). Based on these literatures, we can conclude that the
abundance of lithium becomes depleted with decreasing effective
temperature among stars with spectral types of G to K, while also
showing a wide dispersion for stars with spectral types of F8 to G0.
Such behavior demonstrates that convective mixing is more severe for
G and K type stars. Therefore, giants with spectral types of G and K
are expected to show lower lithium abundances.

The lithium behavior as a function of stellar rotation (e.g. De
Medeiros et al. 2000, de Laverny et al. 2003, L$\grave{e}$bre et al.
2006) and chromospheric activity (Ghezzi et al. 2010, Takeda et al.
2010) have also been studied by many investigations. It is suggested
that fast rotation and active chromospheric activity lead to high
A$_{\rm{Li}}$, where A$_{\rm{Li}}$ = log$\varepsilon$(Li). However,
the effect of age is related to the effect of rotation and
chromospheric activity, since young stars rotate faster and exhibit
more violent chromospheric activities than old stars.

The lithium behavior in dwarfs with and without planets has been
well studied, but there are conflicting conclusions. Some studies
(e.g. Israelian et al. 2004, Takeda \& Kawanamoto 2005, Chen et al.
2006, Israelian et al. 2009) have found that lithium abundance
dilution is slightly higher in stars with planets compared to those
without them. Also, other studies (e.g. Luck \& Heiter 2006, Baumann
et al. 2010) found that there is no difference in lithium abundance
between stars with and without planets. However, such kind of
researches have not been carried out on a large sample of giants. A
study of lithium abundances derived from a large homogeneous sample
is desired, which allows a better understanding of lithium
depletion. With more and more planets (23 planets up until now)
released from the Okayama Planet Search program (Sato et al. 2003)
and the Xinglong Planet Search program (Liu et al. 2008), this gives
us a chance to systematically study the effect of planets on the
lithium abundance for G/K giants.


As is known, chemical composition is different in thin and thick
disk stars in the solar neighborhood, particularly for the
$\alpha$-elements and oxygen (e.g. Bensby 2005; Reddy et al. 2006).
But whether the lithium abundance is different in thin/thick-disk
stars was not known before Ram\'{\i}rez et al. (2012), who
investigated the behavior of lithium abundance in thin-/thick-disk
stars among dwarfs and subgiants. They observed a difference in
lithium abundance between thin- and thick-disk stars, reflecting
different degrees of lithium depletion rather than differences in
lithium enrichment of the interstellar medium. With a catalog
supplemented from the literature (Luck $\&$Heiter 2007, hereby
Luck07), our sample has been enlarged to 565 giants, which can be
used to explore the different properties of thin-/thick-disk giants.

In this paper, we analyze the lithium abundance for 378 G/K giants.
The purpose of this work is to investigate the behavior of the
lithium abundance as a function of metallicity, effective
temperature, and stellar rotation, as well as to explore if there is
any difference between giants with and without planets. The paper is
organized as following: section 2 gives the sample selection and
observational data; section 3 describes the method of analysis;
section 4 presents our discussion; section 5 confirms the lithium
abundance as a function of metallicity and effective temperature and
the effect of thin-/thick-disk star in a sample of literature
compilation 565 giants; finally the results are summarized in the
last section.

\section{Sample selection and observational data}
The sample stars analyzed here are comprised of 321 giants from the
Okayama Planet Search Program (Sato et al. 2003) and 57 giants from
the Xinglong Planet Search Program (Liu et al. 2008). Both programs
aim to detect planets around intermediate-mass G-type (and early
K-type) giants. The spectra were taken with the HIgh Dispersion
Echelle Spectrograph (HIDES) at Okayama Astrophysical Observatory
(OAO), which was equipped at the coude focus of the 1.88 m telescope
during 2008-2010. Before October 2008, one CCD system with
wavelength coverage of 5000-6200 \AA\ was used, and after that the
new mosaic 3 CCD system with wavelength coverage of 4000-7540 \AA\
replaced it. For all the 378 stars, several spectra were obtained
with the iodine absorbtion cell in the planet search program. In
such cases, two spectra with the highest S/N (all higher than 200
per pixel) were selected for each star. Meanwhile, spectra without
the iodine cell were obtained for 71 stars as a comparison sample.
The S/N of these pure stellar spectra ranges from 100 to 250. A slit
width of 200 $\mu$m was adopted, corresponding to a spectral
resolution of 67,000.

The 321 giants from the Okayama Planet Search Program have been
analyzed by Takeda et al. (2008), who determined the stellar
parameters, iron abundances and rotational velocities. The 57 giants
from the Xinglong Planet Search Program were studied by Liu et al.
(2010), in which they have derived the stellar parameters and iron
abundances. Based on the latter work, it was shown that the stellar
parameters of the sample of Liu et al. (2010) are located in the
same region of the Hertzsprung-Russel(H-R) diagram as those of
Takeda et al. (2008), thus we can consider the 378 stars as a
uniform sample in the following analysis.

\section{Analysis}

The ODF line-blanketed analysis with local thermodynamic equilibrium
(LTE) is performed based on the model atmosphere MAFAGS developed
and discussed by Fuhrmann et al. (1997). The stellar parameters of
our sample are taken from Takeda et al. (2008) and Liu et al.
(2010). Although the methods to determine stellar parameters are
different, the results are quite consistent (see Liu et al. 2010).
We adopt the oscillator strengths from the NIST database, with
$\log{gf}$ = 0.002 and -0.299 for the 6707.76 and 6707.91 \AA\
lines, respectively. The collisional broadening parameters which
describe the van der Waals interaction with the hydrogen atoms are
taken from Barklem et al. (1998).

The lithium abundances are determined using the spectral synthesis
method with the IDL/Fortran SIU software package developed by Reetz
(1993). The synthetic spectrum is represented by a single Gaussian
profile, with the combined broadening effects of stellar rotation,
macroturbulence and the instrument profile. The lithium abundances
are obtained until a best fit is reached between the synthetic and
observed spectra. The contribution of FeI 6707.46 \AA\ line has been
included. Considering that the S/N of all the spectra with the
iodine absorption cell are higher than 200, the noise in the spectra
account for less than 0.5 percent of the continuum spectra. We
therefore set the detection limit as a 3 $\sigma$ error, i.e., 1.5
percent of the continuum. For those stars whose lithium line profile
is not clear enough but stronger than the detection limit, a best
fit value is given and the star is marked as an undistinguished
detection. The best fits obtained between observed and synthetic
spectra for HD116292 (with clear detection of lithium), HD62345
(with undistinguished detection of lithium), and HD4398 (with only
upper limit detected) are shown in Figure 1 as typical cases for our
sample.

\begin{figure}
\epsscale{.50}
\plotone{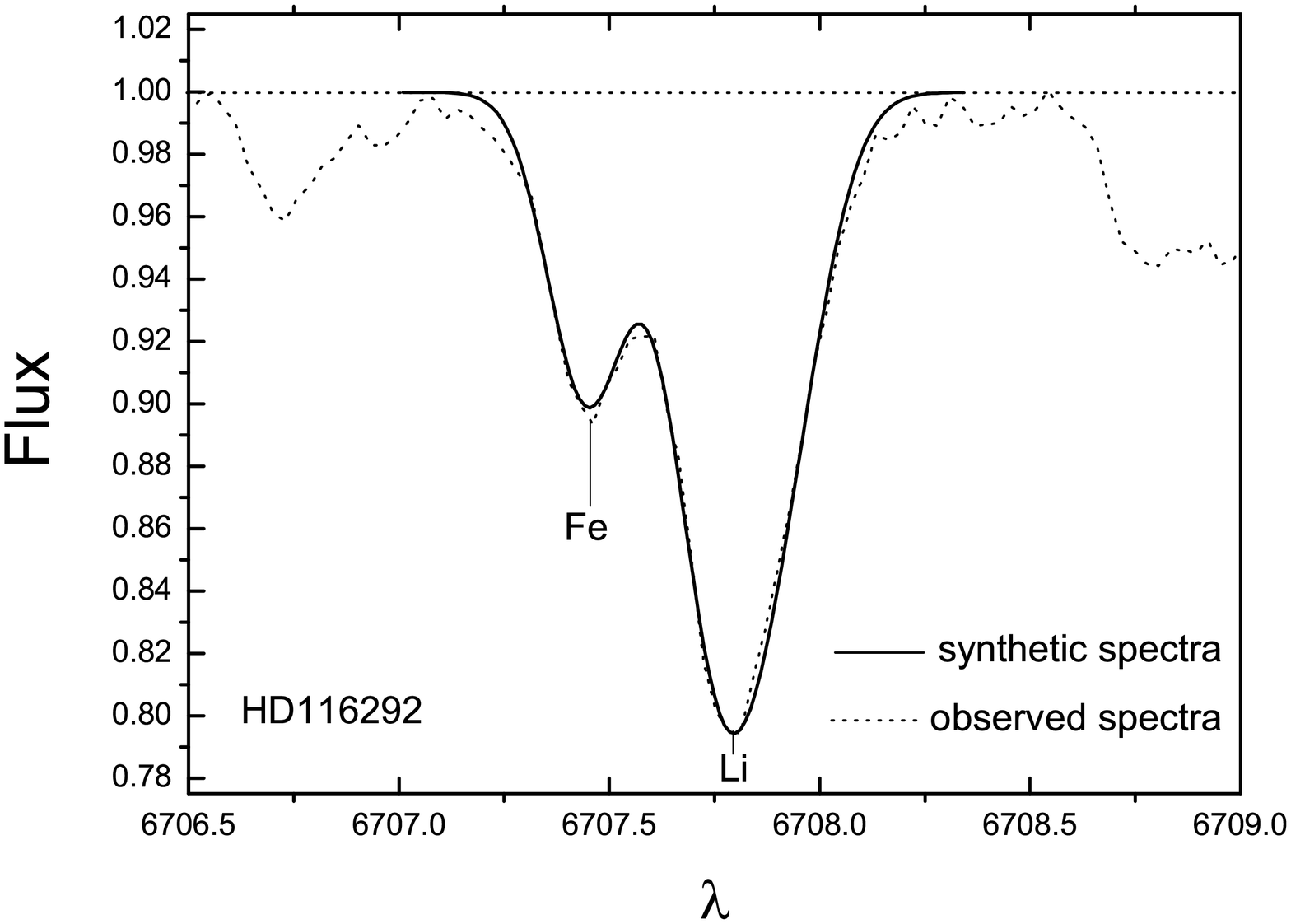}
 \plotone{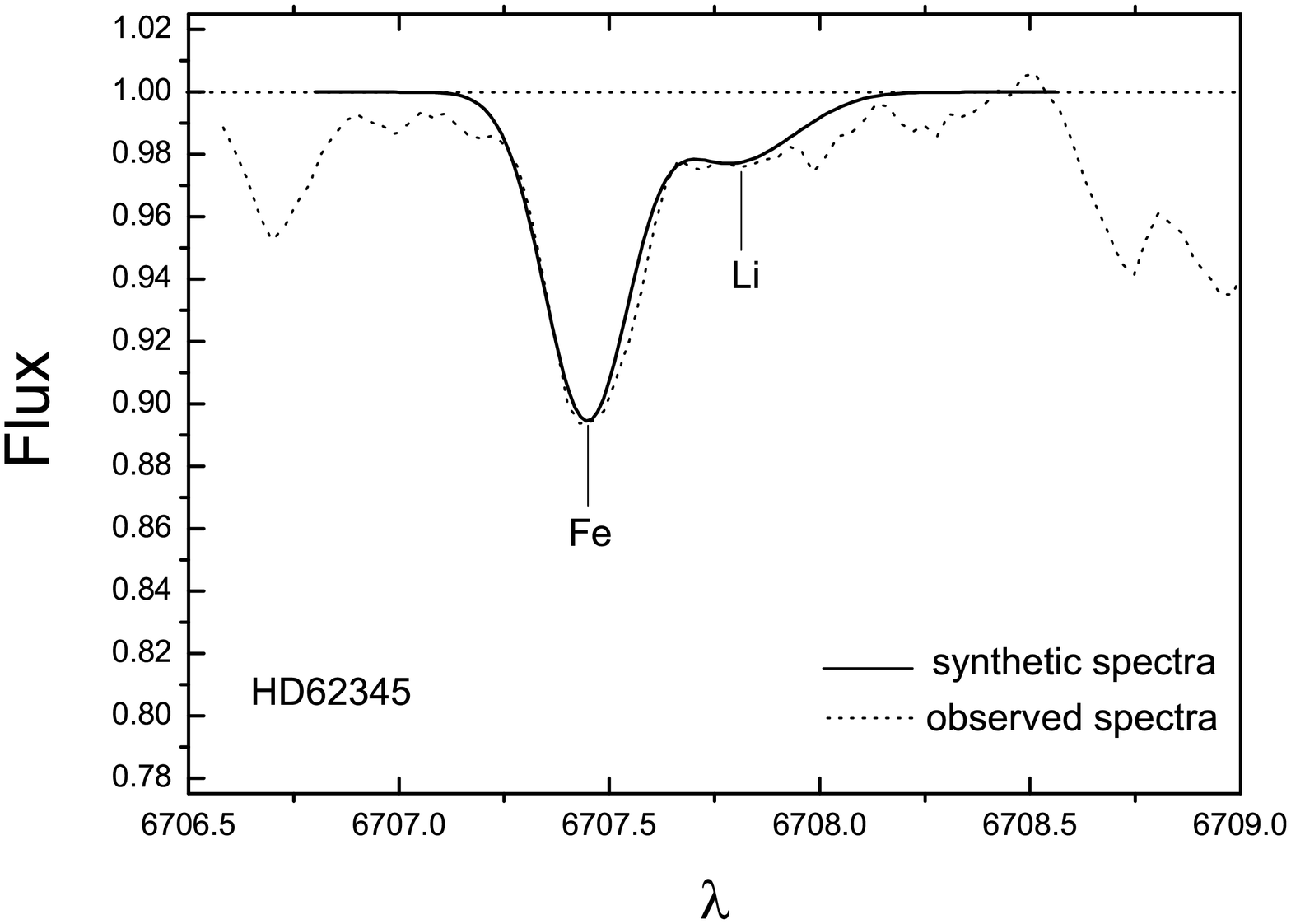}
  \plotone{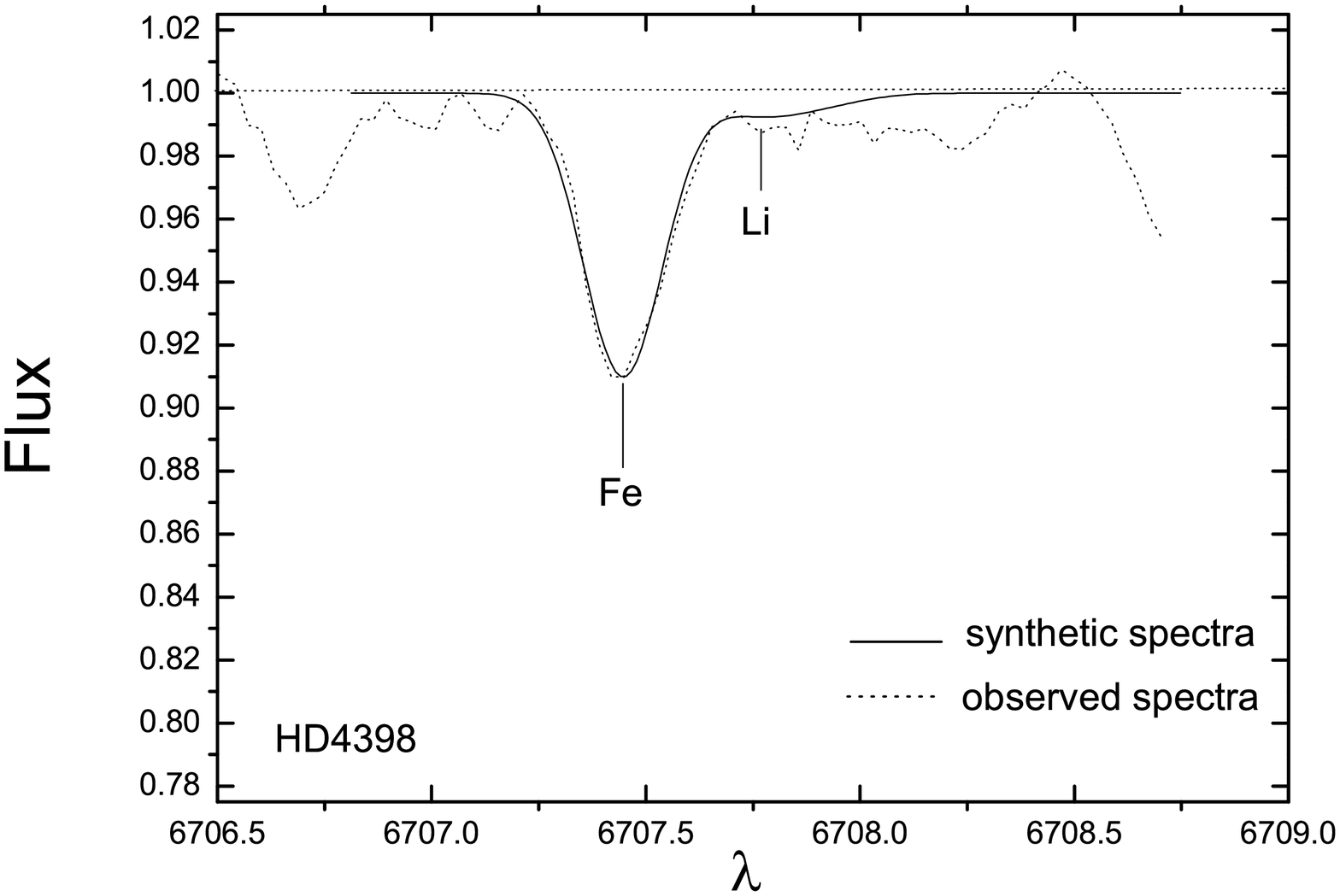}
 \caption{The best fits obtained between observed and synthetic spectra for HD116292, HD62345, and HD4398 (from top to bottom respectively).
\label{Fig. 1}}
\end{figure}
\subsection{Lithium abundance comparison for stars with I$_{2}$ and without I$_{2}$}

For the purpose of precise radial velocity measurements, the spectra
were taken with an iodine absorption cell placed within the optical
beam path, thus the stellar spectral lines located in the region of
5000-6300 \AA\ were heavily blended with the absorption features of
I$_{2}$. The effect on chemical abundance measurement caused by
I$_{2}$ in the spectral region redder than 6300 \AA\ is very weak.
It is shown that the uncertainty of equivalent width caused by
I$_{2}$ around 6700 \AA\ is estimated to be 2 m\AA\ (Wang et al.
2011). We have also obtained both the pure stellar spectra and
I$_{2}$ superposed spectra for 71 stars, and compared the lithium
abundance derived from these two types of spectra to clarify the
pollution of I$_{2}$ in the abundance of lithium for this study.
Table 1 presents the signal to noise ratio and lithium abundance for
the 71 comparison targets determined from both the pure stellar
spectra and I$_{2}$ superposed spectra. Figure 2 compares the
lithium abundances determined from pure stellar spectra and spectra
with I$_{2}$ for the 71 stars. We divide our targets into four
groups: a sub-sample with very clear detections of Li; a sub-sample
with undistinguished detections; a sub-sample with upper limits; and
a sub-sample hosting planets. The four groups are represented by red
open dots, green open squares, blue filled triangles, and black
stars in Figures 2, and 5-11. For the 71 stars, the lithium
abundances can be determined for 44 stars (with clear or
undistinguished detections), and for the other 27 stars, only upper
limits of Li could be derived. The lithium abundances derived from
the spectra with I$_{2}$ are consistent with those derived from the
pure stellar spectra quite well for stars with detected lithium
abundances, giving a difference of 0.00$\pm$0.06 dex, but have a
larger dispersion for stars with upper limit. Since the upper limit
of the lithium abundance of the sub-sample strongly depends on the
quality of the spectrum, namely S/N, the large difference may be due
to the lower S/N of the pure stellar spectra. Therefore, we consider
that the lithium abundances derived from the spectra with I$_{2}$
can represent the true lithium value. It is obvious that, for stars
with clear or undistinguished detection of Li, the lithium
abundances derived from spectra with I$_{2}$ agree well with those
derived from the pure stellar spectra, with a difference of
0.00$\pm$0.06 dex. Meanwhile, the comparison shows a larger
dispersion for stars with upper limits, considering the fact that
the upper limits strongly depend on the quality of the spectra.

\begin{deluxetable}{lrrrr}
\tabletypesize{\small}
 \tablecaption{Comparison of the signal to noise ratio and derived lithium abundance
 based on pure stellar spectra and spectra with I$_{2}$ superposed for 71 stars.
    \label{tbl-1}}
  \tablewidth{0pt}
   \tablehead{
\colhead{Star} & \colhead{S/N$_{\rm{Li(pure)}}$} & \colhead{S/N$_{\rm{Li}}$} & \colhead{A$_{\rm{Li(pure)}}$}   & \colhead{A$_{\rm{Li}}$}\\
} \startdata

HD   9408 & 385 & 305 & $<$-0.06 & $<$-0.01 \\
HD  11949 &  97 & 283 &  0.47 &  0.43 \\
HD  13994 & 222 & 282 &  $<$0.22 &  $<$0.32 \\
HD  15779 & 323 & 336 &  0.23 &  0.38 \\
HD  16400 & 279 & 276 &  0.22 &  0.22 \\
HD  18970 & 320 & 424 &  0.26 &  0.21 \\
HD  22675 & 341 & 263 &     0.62 &     0.57 \\
HD  27371 & 311 & 459 &     1.09 &     1.13 \\
HD  27697 & 340 & 359 &     1.05 &     1.07 \\
HD  28305 & 278 & 424 &     0.80 &     0.85 \\
HD  28307 & 321 & 320 &     1.20 &     1.23 \\
HD  36079 & 242 & 368 &  $<$0.23 &  $<$0.23 \\
HD  39007 & 367 & 249 &     0.36 &     0.46 \\
HD  41361 & 333 & 297 &     1.23 &     1.25 \\
HD  43039 & 325 & 296 &  $<$0.11 & $<$-0.04 \\
HD  45410 & 289 & 291 &  $<$0.10 &  $<$0.20 \\
HD  50522 & 350 & 344 &  0.44 &      0.42 \\
HD  54810 & 274 & 281 & $<$-0.12 & $<$-0.09 \\
HD  57727 & 242 & 339 &     1.34 &     1.34 \\
HD  61363 & 273 & 320 & $<$-0.03 & $<$-0.03 \\
HD  62509 & 354 & 222 &     0.79 &     0.82 \\
HD  71115 & 242 & 305 &     0.78 &     0.83 \\
HD  71369 & 363 & 335 &     0.84 &     0.71 \\
HD  79181 & 218 & 261 &     0.24 &     0.29 \\
HD  81688 & 244 & 321 & $<$-0.11 & $<$-0.11 \\
HD  82210 & 312 & 225 &     1.18 &     1.17 \\
HD  84441 & 357 & 354 &  $<$0.44 &  $<$0.44 \\
HD  91190 & 215 & 351 &     0.55 &     0.62 \\
HD  92125 & 242 & 339 &  $<$0.61 &  $<$0.71 \\
HD  93291 & 194 & 241 &  $<$0.38 &  $<$0.28 \\
HD  94497 & 198 & 321 &  0.16    &  0.24 \\
HD 101484 & 244 & 257 &  0.46 &  0.41 \\
HD 103484 & 172 & 247 &  $<$0.37 &  $<$0.37 \\
HD 104979 & 214 & 361 &  $<$0.13 &  $<$0.23 \\
HD 104985 & 190 & 259 & $<$-0.17 &  $<$-0.07 \\
HD 106057 & 215 & 270 &     0.7 &     0.75 \\
HD 106714 & 233 & 309 &     0.75 &     0.72 \\
HD 107383 & 209 & 382 &     0.48 &     0.55 \\
HD 109317 & 191 & 263 &  $<$0.33 &  $<$0.48 \\
HD 113226 & 207 & 463 &     0.50 &  0.6 \\
HD 120048 & 193 & 224 &     1.39 &     1.44 \\
HD 120420 & 146 & 268 &     0.64 &     0.60 \\
HD 136512 & 319 & 319 &     0.34 &  0.34 \\
HD 136956 & 353 & 258 &     0.66 &  0.64 \\
HD 138905 & 306 & 479 &     1.27 &     1.33 \\
HD 142091 & 320 & 379 &  $<$0.33 &  $<$0.33 \\
HD 146791 & 387 & 370 &  $<$0.36 &  $<$0.41 \\
HD 148604 & 245 & 255 &     1.20 &     1.18 \\
HD 154084 & 314 & 344 &  $<$0.27 &  $<$0.17 \\
HD 167042 & 234 & 302 &  $<$0.23 &  $<$0.28 \\
HD 167768 & 299 & 232 &     0.18 &     0.13 \\
HD 188310 & 339 & 318 &  $<$0.20 &  $<$0.10 \\
HD 188650 & 310 & 324 &     1.01 &     0.91 \\
HD 199665 & 391 & 309 &     0.48 &     0.53 \\
HD 212496 & 305 & 351 &  $<$0.10 & $<$-0.07 \\
HD 219139 & 284 & 287 &  0.19 &  0.19 \\
HD 221345 & 308 & 341 &  $<$0.19 &  $<$0.14 \\
HD   6482 & 263 & 264 &  0.55 &  $<$0.55 \\
HD   6557 & 160 & 278 &  $<$0.11 &  $<$0.11 \\
HD  22799 & 243 & 285 &  0.19 &  0.09 \\
HD  23183 & 195 & 249 &  0.34 &  0.34 \\
HD  41125 & 280 & 251 &  0.59 &  0.54 \\
HD  82638 & 221 & 229 &  $<$0.20 &  $<$0.20 \\
HD 130025 & 316 & 244 &  0.59 &  0.44 \\
HD 136366 & 212 & 199 &     0.57 &     0.62 \\
HD 141853 & 219 & 202 &  $<$0.04 &  $<$0.14 \\
HD 173398 & 323 & 279 &  0.51 &  0.51 \\
HD 173416 & 261 & 326 &  0.22 &  0.12 \\
HD 175679 & 263 & 265 &     0.99 &     0.97 \\
HD 206005 & 257 & 256 &  $<$0.18 &  $<$0.23 \\
HD 220465 & 223 & 265 &  $<$0.22 &  $<$0.07 \\
\hline
\enddata
\end{deluxetable}

\begin{figure}
\epsscale{.60}
 \plotone{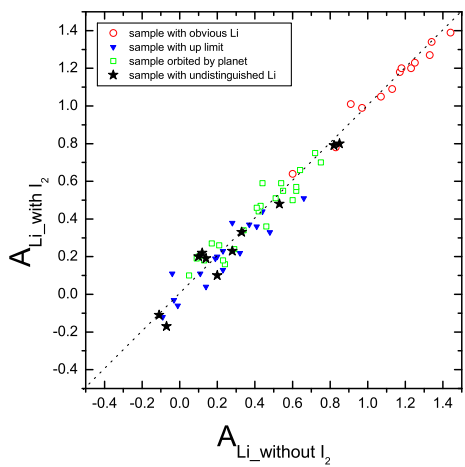}
   \caption{Comparison of the lithium abundances determined from the spectra with and without I$_{2}$ for 71 comparison stars.
   The red open dots, filled squares, open triangles and stars
represent the sub-sample groups with clear Li detection, with
undistinguished Li detection, with Li upper limits and with planets,
respectively. \label{Fig. 2}}
\end{figure}

\subsection{Error analysis}

The error in determining the lithium abundance with the method of
spectrum synthesis mainly comes from two aspects: the systematic
errors introduced by the atmospheric parameters and the uncertainty
in the stellar continuum which could lead to differences in the
lithium abundance.

It is known that the lithium abundance is very sensitive to the
effective temperature, but the effects of surface gravity, iron
abundance and microturbulence velocity are negligible. We calculate
the variance of the lithium abundances due to changes of 100 K in
effective temperature, 0.1 dex in surface gravity, 0.1 dex in
metallicity, and 0.2 km s$^{-1}$ in microturbulence velocity for a
typical star HD12339. These values correspond to the maximum error
of stellar parameters of our sample (see Takeda et al. 2008, Liu et
al. 2010). The typical error of 100 K in effective temperature for
giants will lead to a deviation of 0.1 dex in the lithium abundance,
however the uncertainty of the lithium abundance is negligible by
changes of 0.1 dex in surface gravity, 0.1 dex in metallicity, and
0.2 km s$^{-1}$ in microturbulence velocity. Therefore, the error of
Li abundance due to uncertainties in stellar parameters is at the
level of 0.1 dex.

As for uncertainty in the stellar continuum, since all spectra used
to derive lithium abundance are with S/N higher than 200, the
abundance error caused by location in the continuum is estimated to
be less than 0.5 per cent in the worst case, which results in an
error of lithium abundance of $\pm$0.1 dex at most. The overall
uncertainty in lithium abundance for our sample is less than 0.14
dex.

\subsection{Non-LTE correction}
According to Lind et al. (2009), the non-LTE effect can be very
large on lithium abundance in giants. We perform the non-LTE
calculations of lithium based on the grid and interpolation code by
Lind et al. (2009). For stars whose upper limits of Li abundances
are very small, the correction cannot be determined from the code
because the corresponding equivalent widths are beyond the grids. It
is shown in the grids that the non-LTE correction becomes
independent of A$_{\rm{Li}}$ when it is less than 0.6 dex;
therefore, we take the minimum value from the grid and obtain the
corresponding correction. For a few stars with microturbulence
velocity of 0.9 km s$^{-1}$, which is out of the range of
microturbulence velocity in the grid, the correction corresponding
to 1.0 km s$^{-1}$ is then adopted. Considering that the calculation
for non-LTE correction is related to effective temperature,
metallicity, surface gravity and microturbulence, the non-LTE
corrections are thus plotted against these four parameters in Figure
3. As expected, the non-LTE correction is a function of effective
temperature; the lower the temperature, the higher the correction
value. However, the correction shows no clear correlation with
metallicity, surface gravity or microturbulence velocity. Table 2
presents the star name, effective temperature, surface gravity,
metallicity, stellar mass, lithium abundance or upper limit of the
four sub-sample groups as described in the text, as well as the
non-LTE correction. The range of non-LTE correction is 0.05-0.28
dex, with an average of 0.18 dex for most sample stars. Therefore,
in G and K giants, the non-LTE effect on lithium abundance can not
be ignored.

\begin{deluxetable}{lccrclllll}
\tabletypesize{\small}
 \tablecaption{The stellar parameters and lithium abundances of 378 red giants\label{tbl-2}}
  \tablewidth{0pt}
   \tablehead{
\colhead{Star} & \colhead{T$_{\rm eff}$(K)}   & \colhead{log $g$} & \colhead{[Fe/H]} &\colhead{$M^{*}(\rm{M_{\odot})}$}&\colhead{A1}&\colhead{A2}&\colhead{A3} &\colhead{A4}&\colhead{$\Delta$NLTE} \\
} \startdata

HD87    &5072 &2.63 &-0.07  &2.74  &0.99  &--    &--    &-- & 0.15\\
HD360   &4850 &2.62 &-0.08  &2.34  &--    &0.4   &--    &-- & 0.19\\
HD448   &4780 &2.51 & 0.03  &2.25  &--    &--    &0.36  &-- & 0.22\\
HD587   &4893 &3.08 &-0.09  &1.58  &--    &0.37  &--    &-- & 0.17\\
HD645   &4880 &3.03 & 0.07  &1.95  &0.63  &--    &--    &-- & 0.19\\
..............\\
\hline
\enddata
\tablecomments{Table 2 is published in its entirety in the
electronic edition of The Astrophysical Journal. A portion is shown
here for guidance regarding its form and content.}
\end{deluxetable}

\begin{figure}
\epsscale{1.10}
 \plottwo{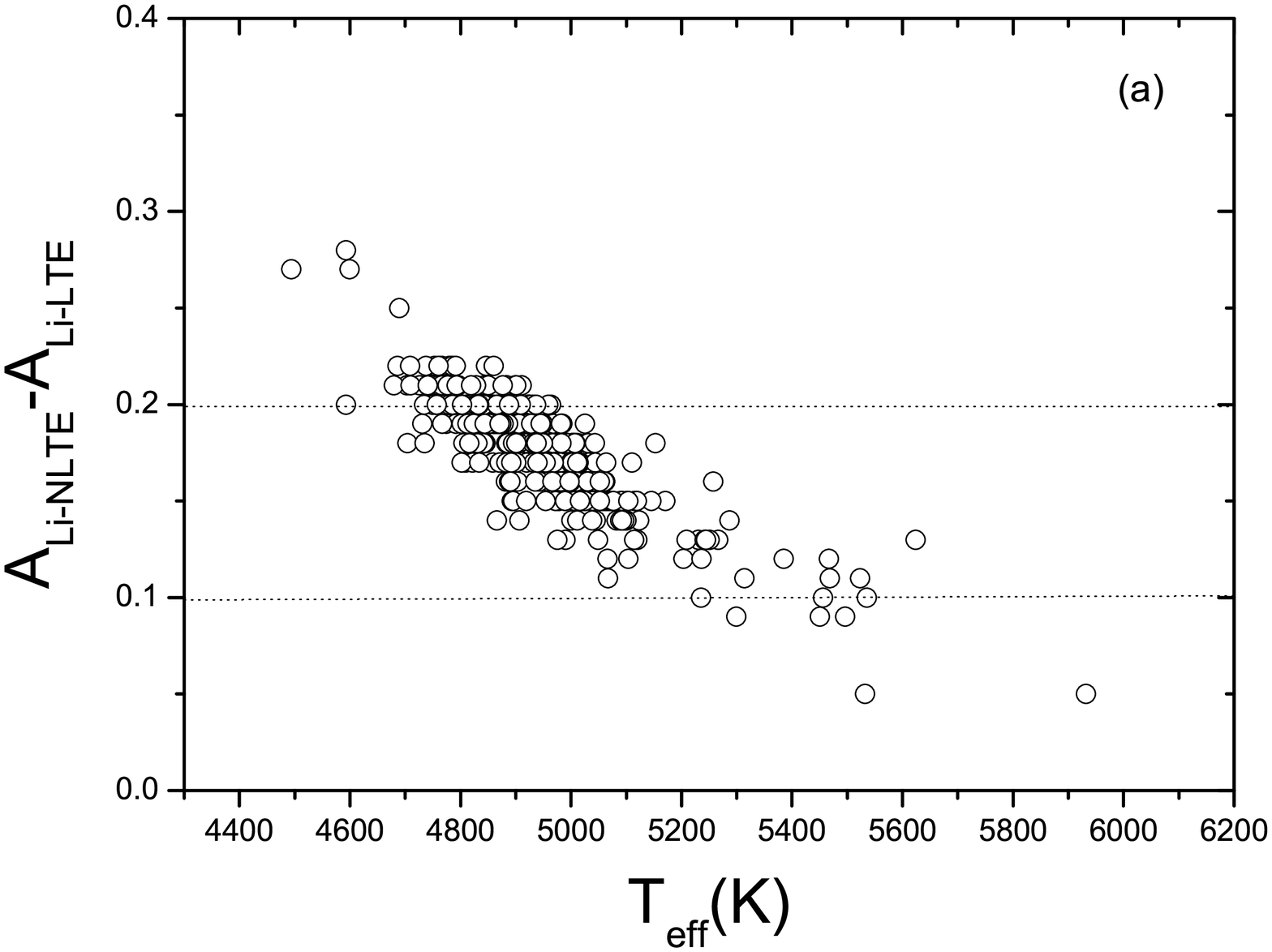}{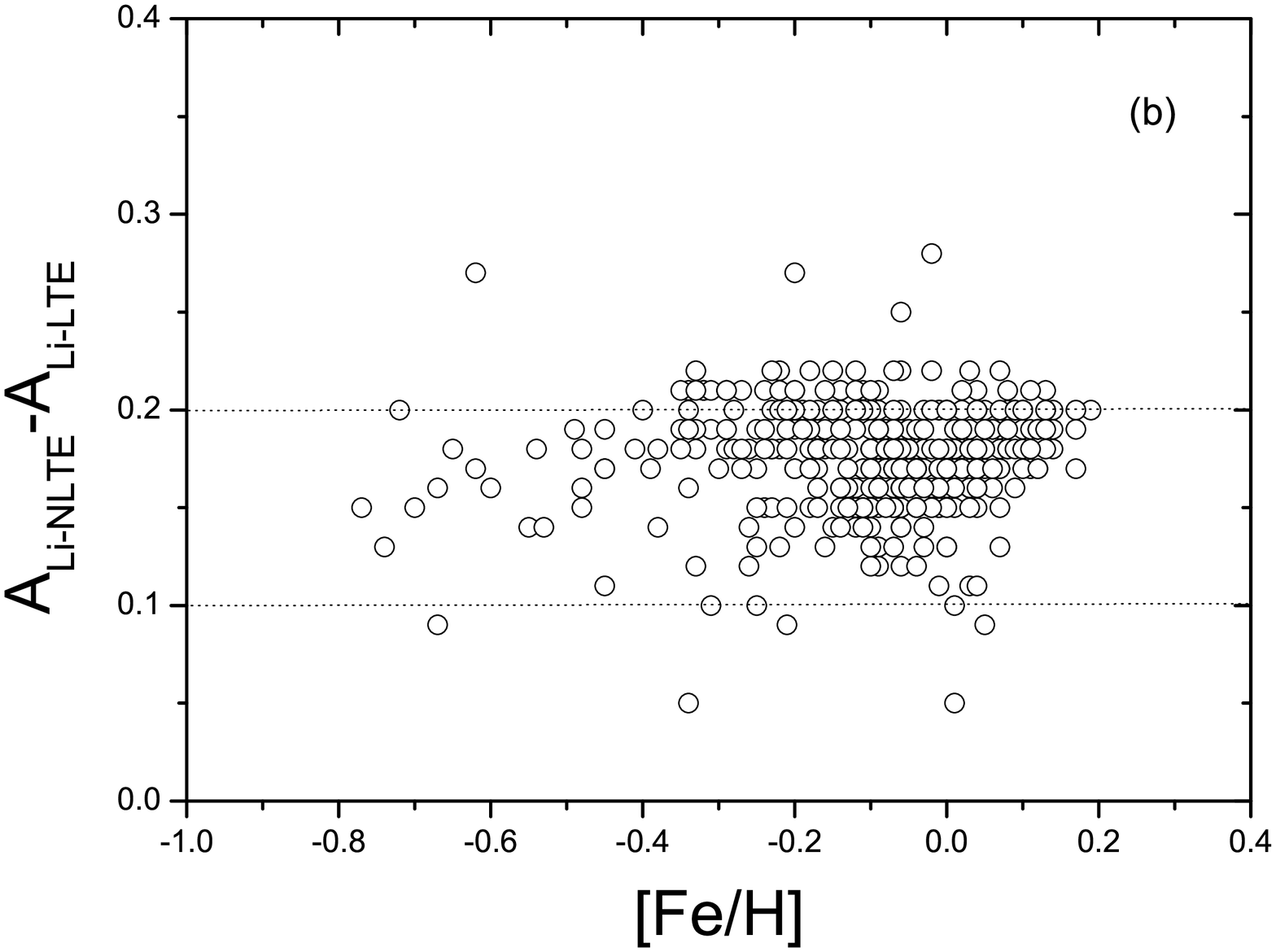}
 \plottwo{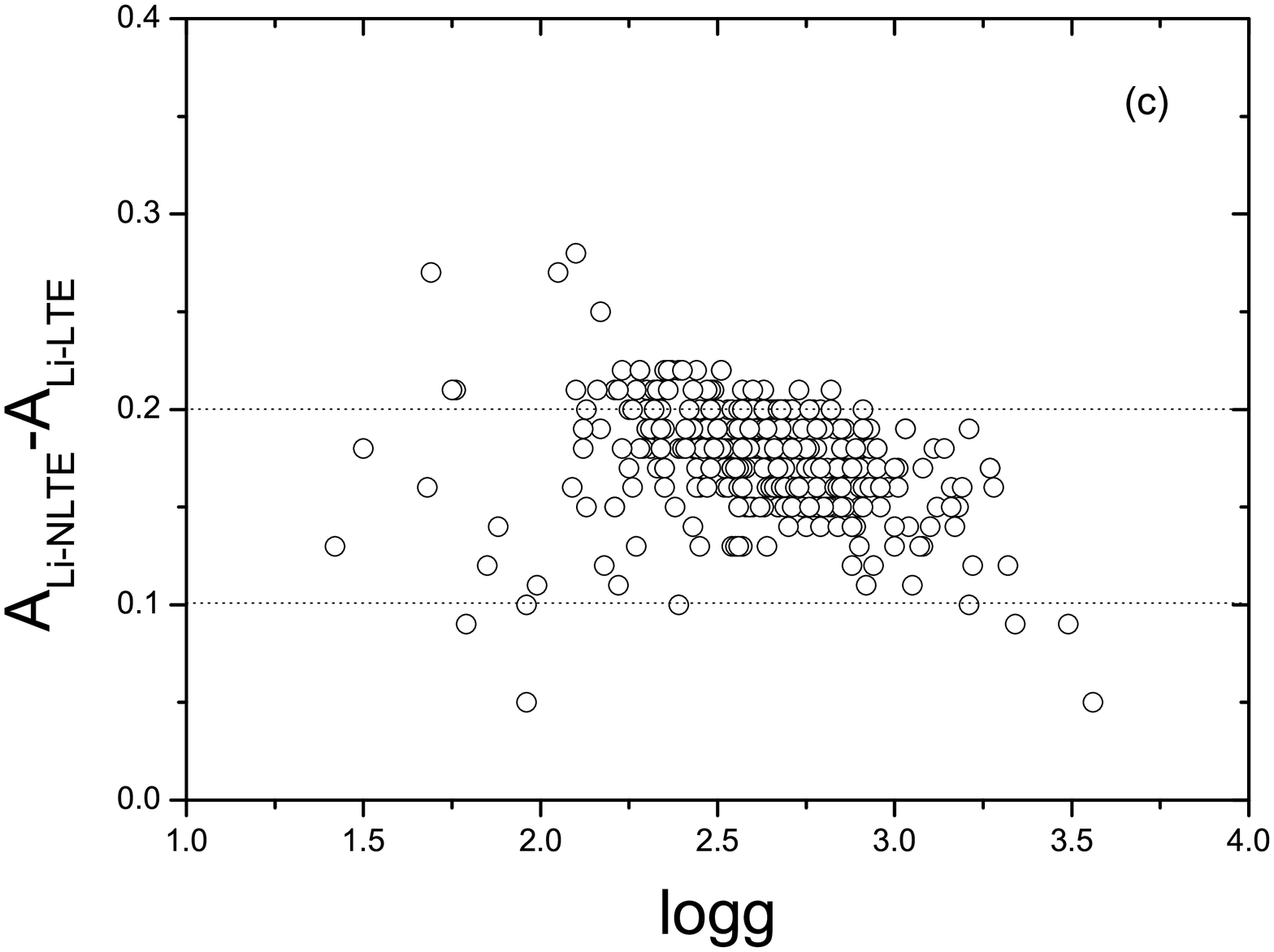}{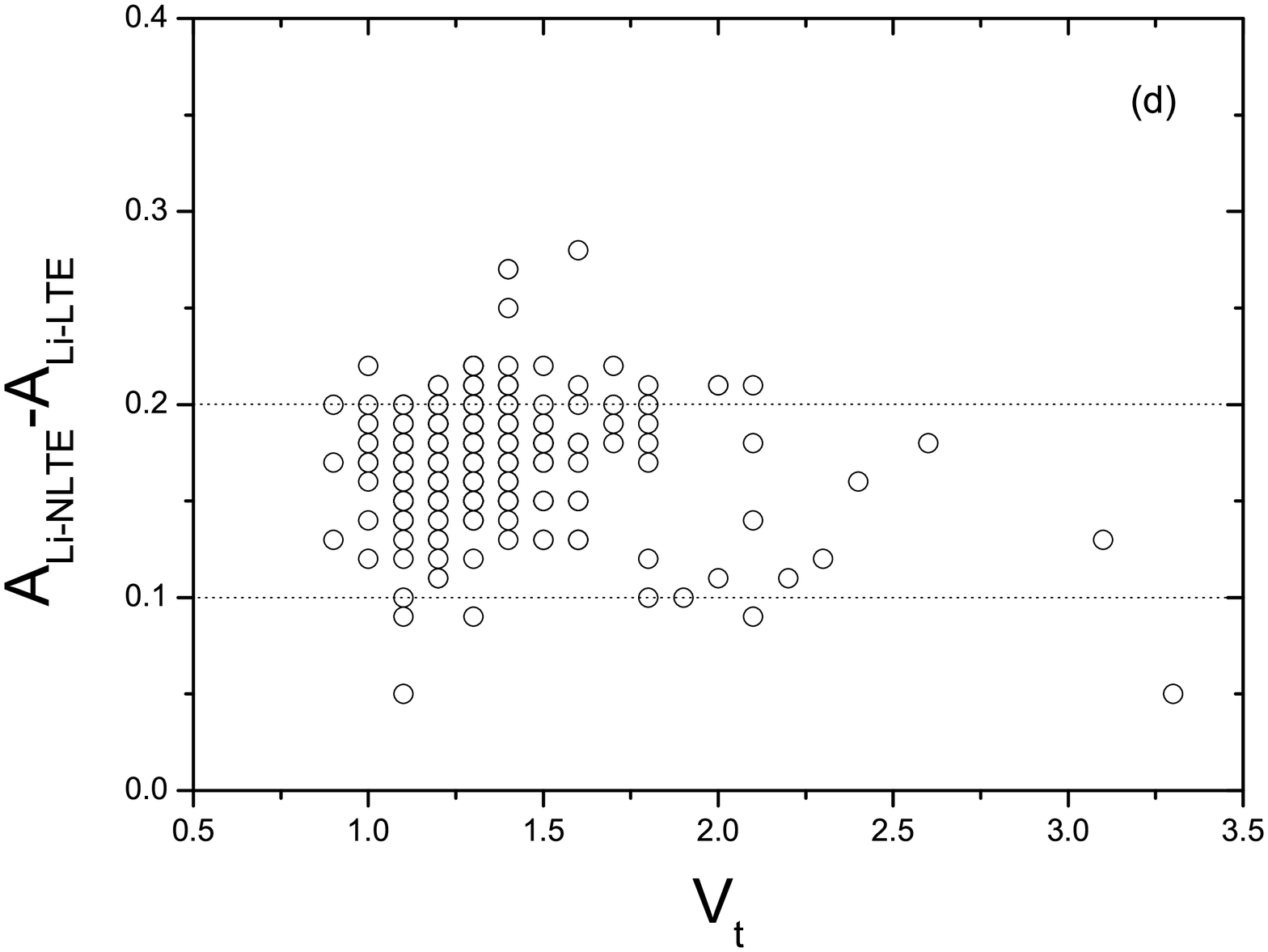}

   \caption{The non-LTE correction for lithium abundance against effective temperature (a), metallicity (b), surface gravity (c),
   and microturbulence velocity (d) for our sample.
    \label{Fig.3}}
\end{figure}

\subsection{Comparison with literatures}

We compare the derived lithium abundance with previous studies by
Luck07, L$\grave{e}$bre et al. (2006) and Luck $\&$ Wepfer (1995).
Since the lithium abundance from these studies were derived based on
LTE model, we also adopt our LTE results for comparison.

 \textbf{\underline{Comparison with Luck07}}

Based on a large sample of giants with high quality spectra, Luck07
has carried out precise parameter determination and abundance
analysis on many elements. Three sets of parameters and
corresponding lithium abundances are determined for 298 nearby
giants, of which 96 stars are in common with our work. The three
sets of parameters include physical parameters, and two sets of
spectroscopic parameters which based on different stellar
atmospheric models (MARCS and MARCS75). Since the lithium abundance
is more sensitive to effective temperature than to other parameters,
we compare the three sets of effective temperature with this study
(hereby referred to Liu2013) to get the best fit of effective
temperature and the corresponding lithium abundances. From Figure
4(a), the effective temperatures from physical data (filled dots)
are consist with ours with the average difference of 6 $\pm$ 79 K
for 96 common stars, and hence, we compare the lithium abundances
based on physical parameters (filled dots) with our results in
Figure 4(b). For the 43 stars with Li detections in both studies,
our lithium abundances are quite close to those of Luck07 with the
average difference being 0.09 $\pm$ 0.13 dex. For stars with upper
limits of A$_{\rm{Li}}$ $<$ 0, the difference is larger. Taking into
consideration of the following two comparisons with other studies,
it is suggested that the negative lithium abundances seem to be
underestimated by Luck07.

\begin{figure}
\epsscale{.50} \plotone{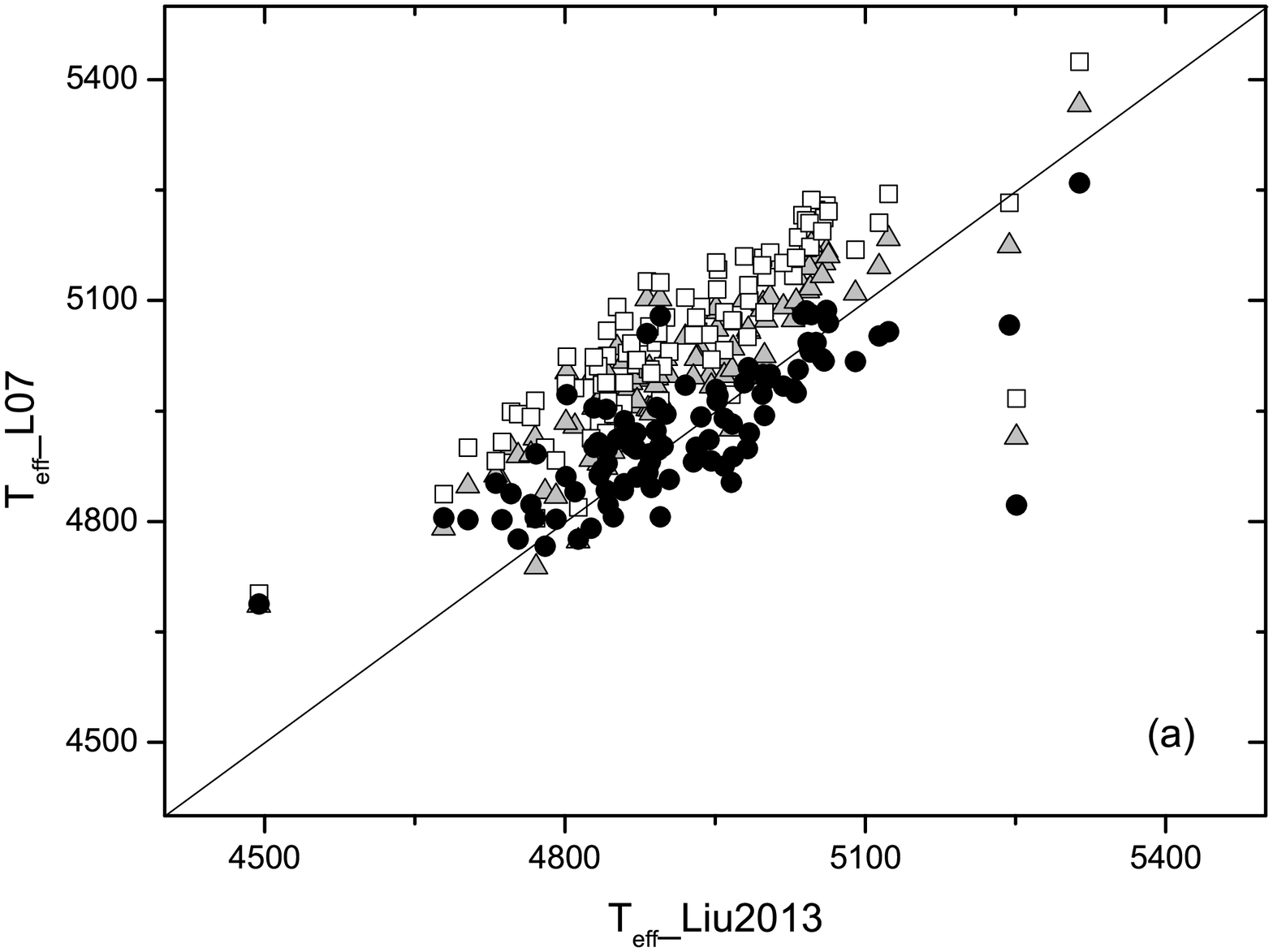} \plotone{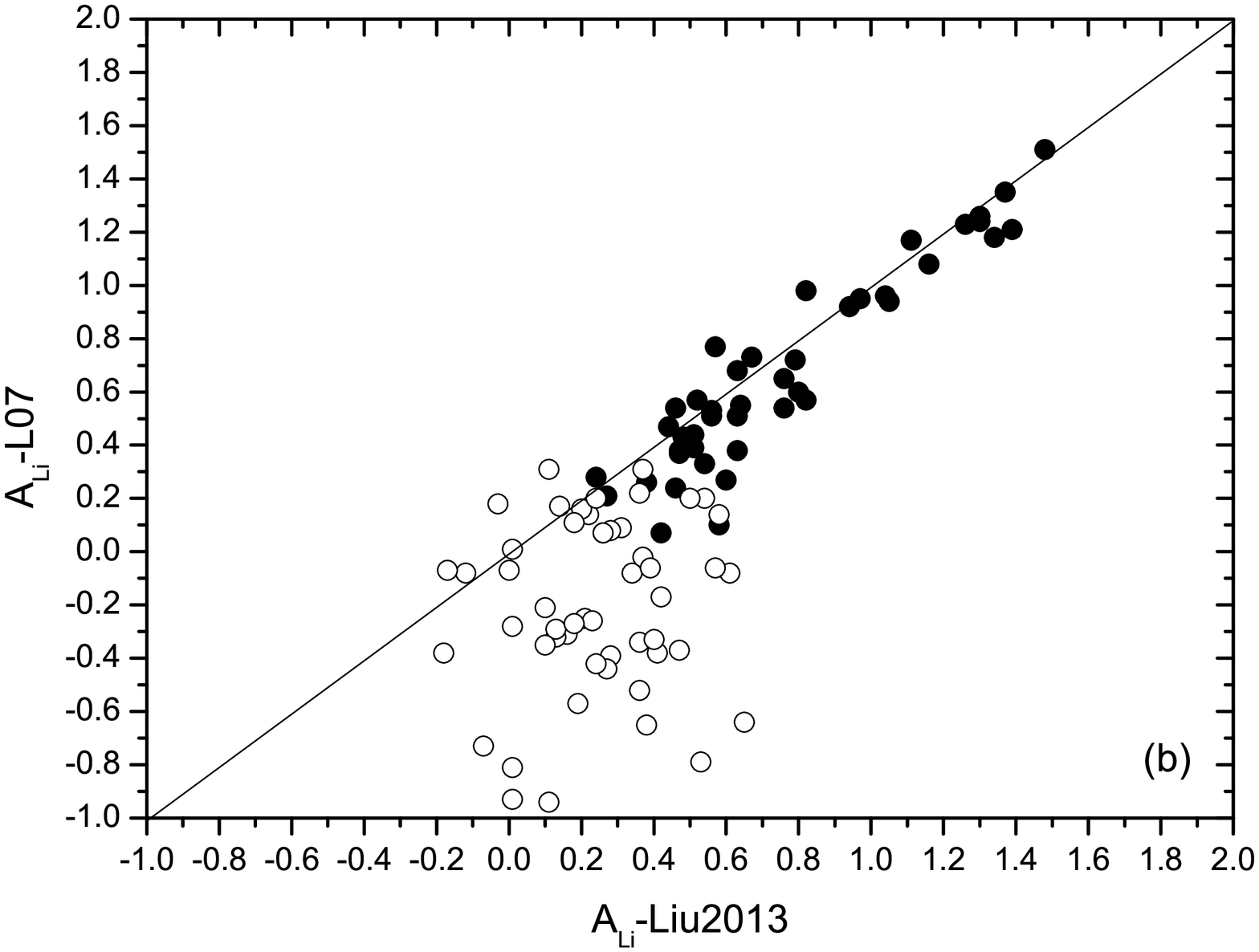}
   \caption{Comparisons of effective temperature and lithium abundances between Luck \& Heiter (2007) and this
   work. Filled dots, triangles and squares represent values from
   physical parameters, and spectroscopic parameters based on two sets of stellar atmospheric models.
    \label{Fig.4}}
\end{figure}

\textbf{\underline{Comparison with L$\grave{e}$bre et al. (2006)}}

Another study on the lithium abundances of a large sample of giant
stars is L$\grave{e}$bre et al. (2006), who derived lithium
abundances for 145 bright giants using the spectrum synthesis
method. Comparisons of effective temperature and lithium abundances
for 17 common stars with our work are presented in Table 3,
including the effective temperature, lithium abundance, metallicity
and the difference of the two works as well. It is noticed that the
lithium abundances are in good agreement, within the measured errors
for all common stars. The differences in effective temperature and
lithium abundance are 53.5 $\pm$ 121K and 0.29 $\pm$ 0.16 dex
respectively; however, the latter is limited to only five stars
showing Li detection in both works. The difference in lithium
abundance for three stars, HD27022, HD65228 and HD76494, is
relatively large, with values of 0.36, 0.37, and 0.46 respectively.
The difference for HD27022 should come from difference in iron
abundance, and that for HD65228 mainly comes from difference in
effective temperature. The large difference for HD76494 remains
unclear; however, the lithium abundance for these three stars agree
with those from Luck $\&$ Wepfer (1995), with the average difference
being less than 0.1 dex. Large differences are also seen for HD71115
and HD159181, which we have referred to undistinguished detection,
while L$\grave{e}$bre et al. (2006) has assigned a uniform upper
limit of 0.4 dex for both of them. Since the spectra from
L$\grave{e}$bre et al. (2006) have much lower S/N (higher than 50)
and lower resolution (45000) than ours, the large difference
probably comes from relatively low quality of spectra in
L$\grave{e}$bre et al. (2006). The lithium abundance of HD71115 from
L$\grave{e}$bre et al. (2006) is remarkably different from that from
Liu2013, but the value from Luck $\&$ Wepfer (1995) agrees very well
with our result.

\begin{deluxetable}{llrrrrrrr}
\tabletypesize{\small}
 \tablecaption{The comparison of effective temperature and lithium
abundances with L$\grave{e}$bre et al. (2006) for 17 common
stars.\label{tbl-3}}
  \tablewidth{0pt}
   \tablehead{
\colhead{Star} & \colhead{T$_{\rm{eff}}$}   & \colhead{T$_{\rm{eff}L06}$} & \colhead{A$_{\rm{Li}}$} &\colhead{A$_{\rm{Li}}$$_{\rm{L06}}$}&\colhead{[Fe/H]}&\colhead{[Fe/H]$_{\rm{L06}}$}&\colhead{$\Delta$T$_{\mathrm{eff}}$} &\colhead{$\Delta$A$_{\rm{Li}}$} \\
} \startdata
HD 1367 &  4982& 5150& $<$0.57 & $<$0.4 & -0.01&0.00 & -168 &  0.17\\
HD 27022&  5314& 5280&  1.26   & 0.9  & -0.01 &-0.20&   34  &  0.36 \\
HD 40801&  4834& 5000& 0.46    &$<$0.4 &  -0.17&-0.30&  -166&  0.06  \\
HD 65228&  5932& 5600&  2.47   & 2.1 & 0.01 &0.0  &   332&  0.37 \\
HD 67447&  4974& 5000&  1.3    &1.2  & -0.06&0.00  &  -26 &   0.1 \\
HD 71115&  5062& 5150&  0.78   &$<$0.4 & -0.07&-0.10  & -88 & $>$0.38 \\
HD 76219&  4904& 5000&  0.43   &$<$0.4 & -0.15&-0.20  &  -96& 0.03   \\
HD 76494&  4828& 4950&  0.86   & 0.4 & -0.20&-0.30 &  -122&   0.46\\
HD 98839&  4936& 5101&  -      &$<$0.4&  -0.05&0.00      &    -74 &- \\
HD 106057& 4956& 5100& $<$0.76 &$<$0.4&  -0.10&-0.20&   -144&   0.36 \\
HD 109379& 5145& 5150&  1.06   & 0.9&  -0.01&-0.3&    -5  &  0.16 \\
HD 119035& 4816& 4900& $<$0.28 & $<$0.4& -0.35&-0.2 &  -84 &   -0.12 \\
HD 150030& 4850& 4900& $<$0.39 &$<$0.4 & -0.09&-0.20&  -50  & -0.01 \\
HD 159181& 5153& 5300&  0.73   &$<$0.4& -0.15&0.30 &  -147& $>$0.33 \\
HD 174980& 5008& 5150& $<$0.53 &$<$0.4&   0.10&0.10  & -142 &   0.13\\
HD 206731& 4936& 4970& $<$0.39 &$<$0.4&  -0.14&-0.3  & -34 &   -0.01 \\
HD 210807& 5071& 5000&  0.63 & $<$0.4& -0.10&-0.10 &  71  &   $>$0.23\\
\hline
\enddata
\end{deluxetable}

\textbf{\underline{Comparison with Luck $\&$ Wepfer (1995)}}\\
Luck $\&$ Wepfer (1995) have determined the stellar parameters and
lithium abundances for 38 F/G bright field giants, in which there
are 15 stars in common with our sample. Among these 15 stars, 9 have
clear Li detections in both works. The differences in effective
temperature and lithium abundance are 11.5 $\pm$ 71 K (for 15 stars)
and 0.04 $\pm$ 0.10 dex (for 9 stars) respectively, which indicates
a good agreement between these two studies. Comparisons of the above
measurements are presented in Table 4. Note that the differences of
HD36079 and HD92125 are quite large; they are assigned upper limits
in our work, but with detected abundances from Luck $\&$ Wepfer
(1995). The reason still remains unclear, although we have carefully
checked the spectra and parameters.

\begin{deluxetable}{llrrrrr}
\tabletypesize{\small}
 \tablecaption{The comparison of effective temperature and lithium
abundances with Luck $\&$ Wepfer (1995) for 15 common
stars.\label{tbl-4}}
  \tablewidth{0pt}
   \tablehead{
\colhead{Star} & \colhead{T$_{\rm{eff}}$}   & \colhead{T$_{\rm{eff}(L95)}$} & \colhead{A$_{\rm{Li}}$} &\colhead{A$_{\rm{Li}}$$_{\rm{(L95)}}$}&\colhead{$\Delta$T$_{\rm{eff}}$} &\colhead{$\Delta$A$_{\rm{Li}}$} \\
} \startdata
HD 27022    &5314  &5275 & 1.26 & 1.25    &   39 &  0.01 \\
HD 36079    &5209  &5225 & $<$0.23& 0.68    &  -16 &$>$-0.45 \\
HD 65228    &5932  &5900 & 2.47 & 2.52    &   32 & -0.05 \\
HD 67447    &4974  &4825 & 1.3  & 1.15    &  -51 &  0.15 \\
HD 71115    &5062  &5050 & 0.78 & 0.80    &   12 & -0.02 \\
HD 76219    &4904  &4875 &0.43 & 0.31    &   29 &   0.12 \\
HD 76294    &4844  &4850 & 0.57 & 0.40    &  -6  &  0.17 \\
HD 76494    &4828  &4900 & 0.86 & 1.00    &  -72 & -0.14 \\
HD 84441    &5385  &5300 &$<$0.44 & 0.67    &   85 &$>$-0.27 \\
HD 92125    &5468  &5600 & $<$0.61& 1.01    & -132 &$>$-0.4  \\
HD 99648    &5002  &4850 & 0.42  & 0.32    &  152 & 0.1 \\
HD 109379   &5145  &5125 & 1.06 & 1.03    &   20 &  0.03 \\
HD 144608   &5266  &5200 &$<$ 0.43 & 0.31    &   66 & $<$0.12 \\
HD 150030   &4850  &4775 &$<$0.39 & 0.26    &   75 & $<$0.13 \\
HD 214567   &4989  &5050 &$<$0.27 &-0.26    &  -61 & $<$0.53 \\
\enddata
\end{deluxetable}

\section{Results and discussion}

The lithium abundance is supposed to depend on metallicity,
effective temperature, stellar mass, age, stellar rotation and
chromospheric activity for giants. However, the effect varies from
different samples and different studies. In this work, we study the
behavior of lithium abundance as a function of [Fe/H],
T$_{\rm{eff}}$, and stellar rotation, as well as whether stars that
host planets can affect their lithium abundances. We know that there
are large uncertainties in stellar mass and age for red giants and
red clump giants from the evolutionary tracks, hence we do not
investigate the relations of lithium abundance against stellar mass
and age in this study.

\subsection{Lithium abundance versus [Fe/H]}

The lithium abundance versus metallicity for our sample is plotted
in Figure 5. The open dots, open squares, open triangles and stars
respectively represent the sub-sample with clear Li detections, with
undistinguished Li detections, with upper limits of lithium
abundance, and with planets. The metallicity of our sample ranges
from -0.8 to 0.2, but in the lower-metallicity region with [Fe/H]
$<$ -0.4, only a few stars show Li detections. However, for those
with [Fe/H] $>$ -0.4, about half of them have Li detections.

According to theoretical models, the surface lithium is diluted by a
factor of 28 to 61 for Red Giant Branch (RGB) stars with a mass of
1-5 M$_{\odot}$ (Iben 1965, 1966, 1967a,b). These depletions
correspond to a decrease in lithium abundance of 1.8-1.4 dex
relative to the assumed initial value from the main sequence.
However, some works based on observation found that the depletion is
more severe than the theoretical prediction. For instance, Brown et
al. (1989) derived an average lithium abundance lower than 0.0 dex
for 891 late-G to K giants, which is at least 1.5 dex below the
predicted value; Luck $\&$ Wepfer (1995) found that lithium is
depleted by a factor of 100 to 1000 for bright giants; and de
Laverny et al. (2003) found a dilution by a factor of at least 600
for G0III type giants with a mass of 2 to 3 M$_{\odot}$. In our
sample, Li depletion reached a factor of 770 for objects with Li
detections compared with the current interstellar medium value of
A$_{\rm{Li}}$ $\approx$ 3.3, which is far more diluted than the
theoretical prediction. The over depletion was also found by other
studies focusing on giants (eg. Brown et al. 1989, De Medeiros et
al. 2000, de Laverny et al. 2003). If the over depletion is due to
an extra-mixing mechanism existed in red giants/red clump giants,
the [C/Fe] ratio, which indicates the degree of evolution-induced
envelope mixing, would also reflect this difference; therefore, a
correlation may be expected between A$_{\rm{Li}}$ and [C/Fe], which
is not shown in Figure 6. From literatures we know that early-F type
stars($\sim$ 1.5 M$_{\odot}$) show a considerable diversity in
A$_{\rm{Li}}$ (e.g. Boesgaard \& Tripicco 1986). For A type
stars($\sim$ 2 M$_{\odot}$), A$_{\rm{Li}}$ is ranging from normal to
undetected value (e.g. Takeda et al. 2012). For rapidly rotating A-B
type stars (2-4 M$_{\odot}$), it is very hard to detect the surface
Li value. This is probably because in early type stars, surface Li
is depleted because of the envelope mixing due to meridional
circulation. It is more convincing that Li deficiency may have
already existed in the main sequence phase for many 1.5-5
M$_{\odot}$ stars and was carried over to the evolved G/K giants.

Except for three Li-rich stars with A$_{\rm{Li}}$ $>$ 1.7, the
lithium abundance is increasing towards a higher metallicity in
Figure 5. The dispersion of Li abundance at higher metallicities is
very large (with a maximum of about 2.0 dex).


\begin{figure}
 \plotone{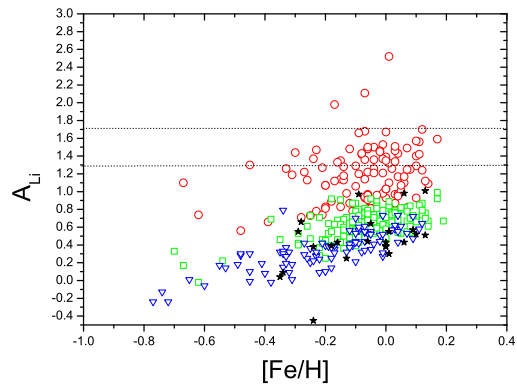}
 \epsscale{.70}
   \caption{Lithium abundance versus metallicity. The symbols
   are the same as in Figure2. Two dash lines divide the regions corresponding to Li-rich,
   Li-normal and Li-depletion.}
\label{Fig. 5}
\end{figure}


\begin{figure}
 \plotone{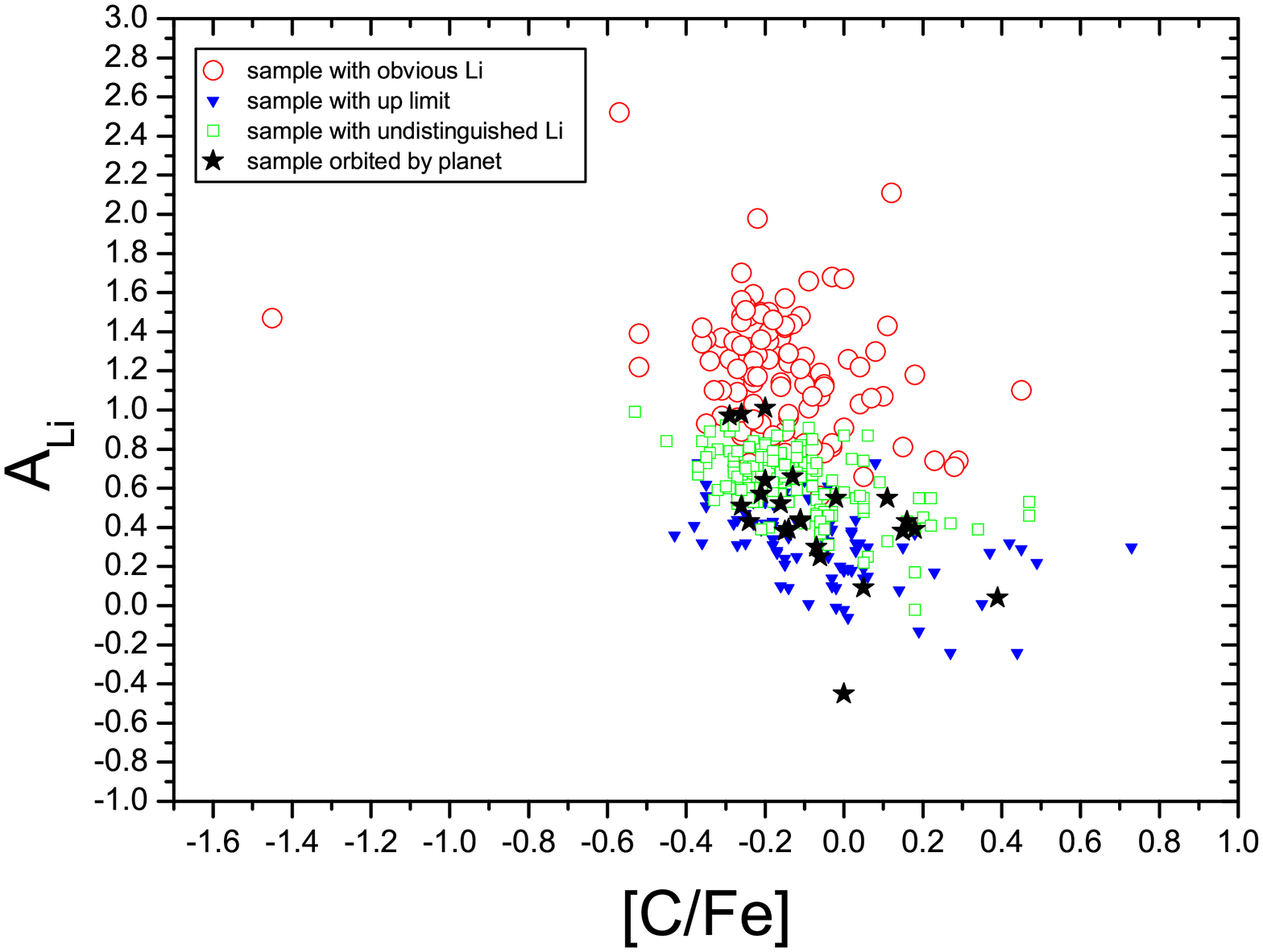}
 \epsscale{.60}
   \caption{Lithium abundance versus [C/Fe]. The symbols
   are the same as in Figure2.} \label{Fig. 6}
\end{figure}

\subsection{Li-rich stars and Li-normal stars}

As predicted by theoretical models, typical RGB stars, with solar
metallicity and stellar mass between 1.0 to 5.0 M$_{\odot}$ and
which have undergone the first dredge up stage, are expected to show
A$_{\rm{Li}}$ $<$ 1.8-1.4 dex. We have chosen the value of
A$_{\rm{Li}}$ $>$ 1.7 dex, and 1.3-1.7 dex as the standard to define
a star as being abnormally Li-rich or Li-normal, as suggested by
some investigations (e.g. Brown et al. 1989, Gonzalez et al. 2009).
Among all samples, the three stars HD65228, HD212430 and HD102845
show A$_{\rm{Li}}$ $>$ 1.7, and are thus classified as Li-rich
giants. There are 36 stars that exhibit the expected lithium
abundances (1.3 $\leq$ A$_{\rm{Li}}$ $\leq$ 1.7), i.e. Li-normal
stars, and the remaining Li-depletion stars show an abundance of
A$_{\rm{Li}}$ $<$ 1.3. Some researches (e.g. Brown et al. 1989,
Charbonnel $\&$ Balachandran 2000, Ram\'{\i}rez et al. 2012) show
that the previous results indicate the fraction of Li-rich giants is
less than 1$\%$, which agrees well with our results. To check the
mass and evolutionary status of the Li-rich and Li-normal giants, we
plot their positions in H-R diagram in Figure 7, by adopting the
stellar evolutionary tracks of YY (Yi et al. 2003) with the mean
metallicity Z = 0.008 ([Fe/H] = -0.4, blue line) and solar
metallicity Z = 0.02 ([Fe/H] = 0, green line). The chosen mass range
is 1.0 to 5 M$_{\odot}$, but most of them lie within 2-4
M$_{\odot}$. Three sizes of the symbols correspond to Li-rich,
Li-normal and Li-depletion giants. Li-rich giants with A$_{\rm{Li}}$
$>$ 1.7 are represented by the largest, filled symbol, and 36
Li-normal giants with 1.3 $\leq$ A$_{\rm{Li}}$ $\leq$ 1.7 by
mid-size symbol. As a result of selection strategy in our sample,
i.e., only late G and K giants are selected for planet search
program, the majority of stars which are classified as RGBs are
located in the same region of the H-R diagram. According to the
statement of Charbonnel \& Balachandran(2000), the Li production
phase is very short, which supports the result of observations that
less than 1$\%$ giants are Li-rich, as confirmed both in previous
studies and our work.

HD65228, possessing the highest lithium abundance in our sample, is
located on the blue side of the Hertzsprung gap. This star just
began to develop a convective envelope, and a high lithium abundance
is expected by the standard theory prediction. For HD212430 and
HD102845, they are located in the clump region of the H-R diagram,
hence there are two possible scenarios to explain such high Li
abundances (Brown et al. 1989). One is that these two stars might be
the first ascent giants, which are partly evolved, and assumed to
have left the main sequence with their initial surface Li
(A$_{\rm{Li}}$ $\approx$ 3.3). The other explanation is that they
have a special circumstance that show high surface Li abundances,
which is called the Cameron-Fowler mechanism (Cameron \& Fowler
1971), and predicts that carbon isotopic ratios will decrease. Li
production precedes the extra-mixing phase which connects the
convective envelope with the CN-burning region to produce the low
$^{12}$C/$^{13}$C ratio. The mixing result in the production of
$^{7}$Li, which is then convected to the surface. We know that 96\%
of the evolved stars show a low $^{12}$C/$^{13}$C ratio (Charbonnel
\& Do Nascimento 1998), which is in disagreement with standard
predictions. Thus the most reasonable explanation is that they have
undergone a period of Li-production. The 36 giants that have normal
amounts of Li, which contain about 10 \% of our sample, are supposed
to show the present interstellar medium abundance of A$_{\rm{Li}}$
$\approx$ 3.3, and have experienced the expected dilution phase
during the first dredge-up phase. The surface lithium abundance
decreases with respect to its value at the end of the main sequence
by a factor of 30 to 60, depending on the stellar mass and
metallicity.

\begin{figure}
\epsscale{.60}
 \plotone{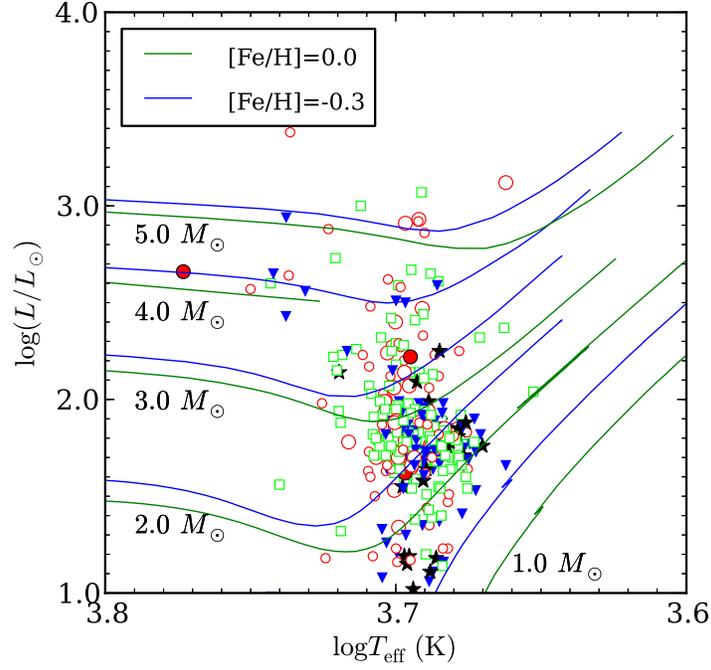}
   \caption{Lithium abundance in H-R diagram. The symbols
   are the same as in Figure2. Three sizes of the
   symbols correspond to Li-rich, Li-normal and Li-depletion giants.\label{Fig.
   7}}
\end{figure}

\subsection{Lithium abundance versus effective temperature}

Figure 8 shows that the lithium abundance is a function of
temperature£¬ with the detection limit increasing when the effective
temperature increases. Generally, from our sample, the lithium
abundance increases towards a higher metallicity for our sample, but
in the narrow range of 4800-5100 K, there is no clear correlation
between lithium abundance and effective temperature. Previous
studies (Wallerstein et al. 1994, De Medeiros et al. 2000, de
Laverny et al. 2003) have demonstrated that the lithium abundance is
a function of effective temperature. However, we also notice that,
within the color index range of 0.8 $<$ B-V $<$ 1.0, the well
established gradual decline of lithium abundances as a function of
effective temperature is not clear in many studies (Wallerstein et
al. 1994, De Medeiros et al. 2000). The lithium abundances span two
orders at a given effective temperature for our sample, suggesting
that the stellar evolution stage and mass vary with dilution (De
Medeiros et al. 2000).

\begin{figure}
\epsscale{.60}
 \plotone{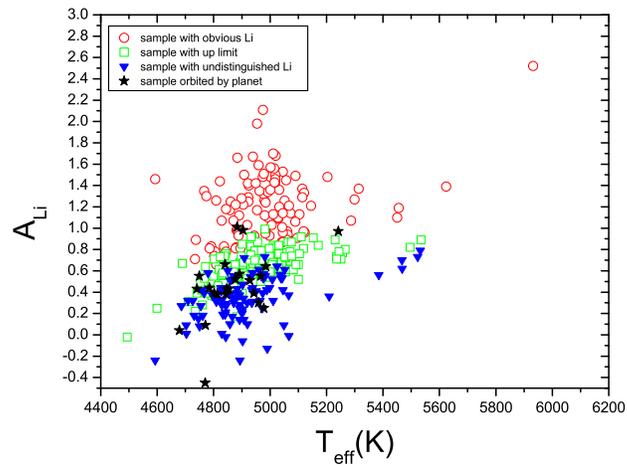}
  \caption{Lithium abundance versus effective temperature. The symbols
   are the same as in Figure2.}
 \label{Fig. 8}
\end{figure}

To get rid of the effect of population, we divide our sample into 3
groups by metallicity: -0.8 $<$ [Fe/H] $\leq$ -0.4,
 -0.4 $<$ [Fe/H] $\leq$ -0.1, and -0.1 $<$ [Fe/H] $\leq$ 0.2; we then plot the lithium
abundance versus effective temperature in Figure 9. To get the
general behavior of lithium in red giants, the three Li-rich stars
(A$_{\rm{Li}}$ $>$ 1.7) HD65228, HD212430 and HD102845 are not
plotted in these figures. It is clear that there is no correlation
between the lithium abundance and effective temperature in the
metallicity range of -0.1 $<$ [Fe/H] $\leq$ 0.2, but with weak
correlation in the metallicity range of -0.4 $<$ [Fe/H] $\leq$ -0.1.
Although there are only four stars with clear Li detections in the
metallicity range of -0.8 $<$ [Fe/H] $\leq$ -0.4, the lithium
abundance as a function of effective temperature is quite clear.

\begin{figure}
\epsscale{.50} \plotone{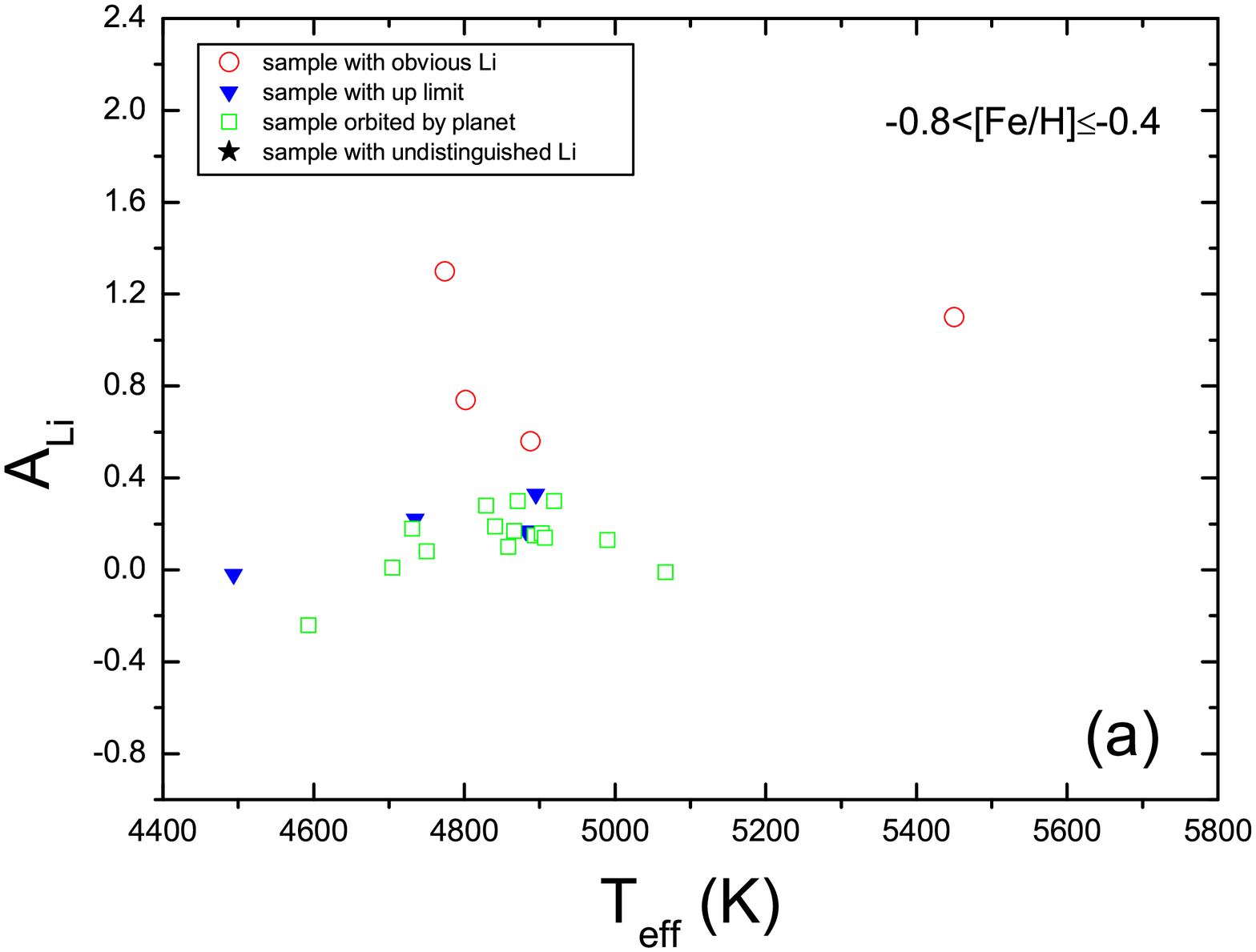} \plotone{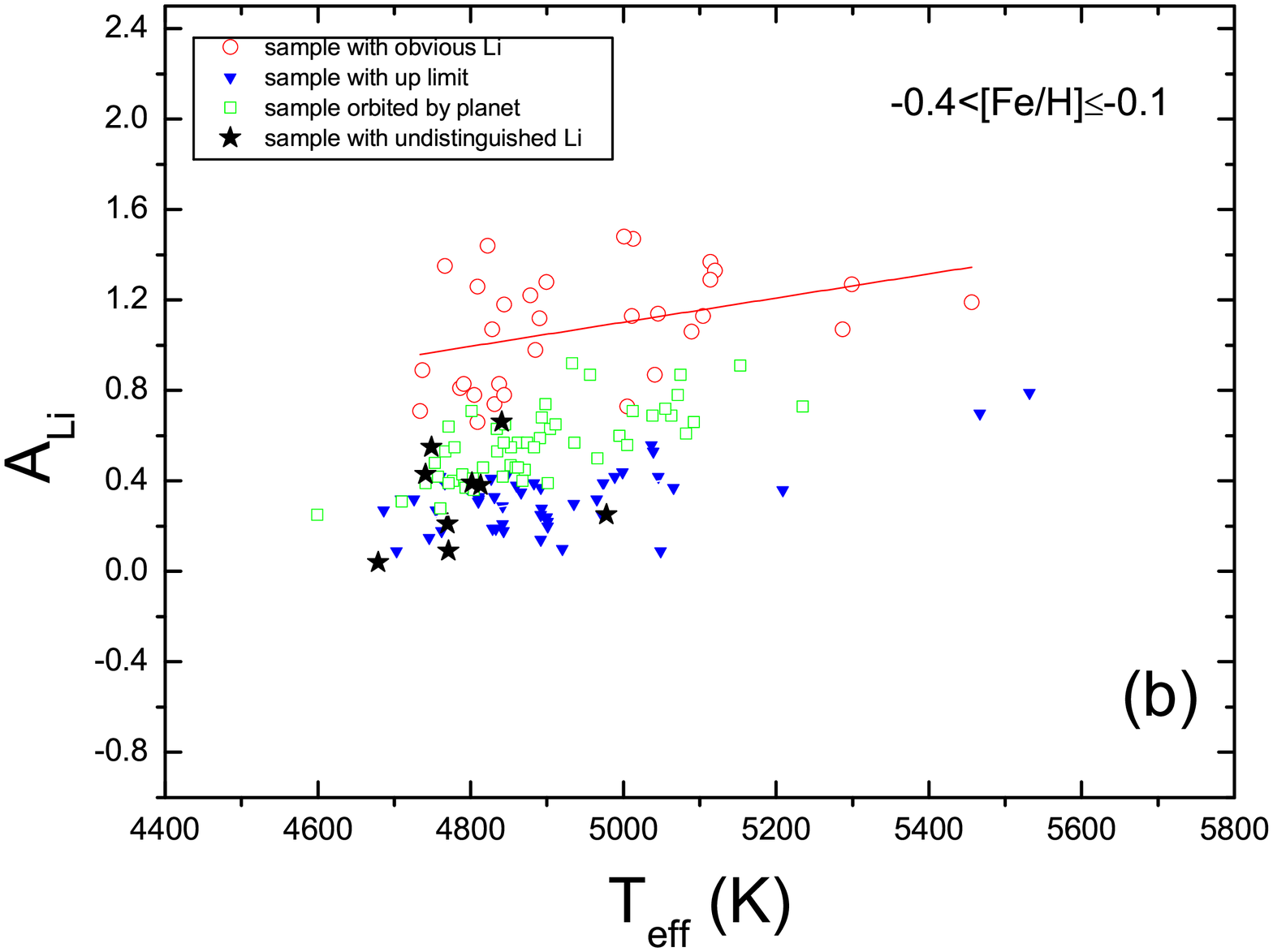} \plotone{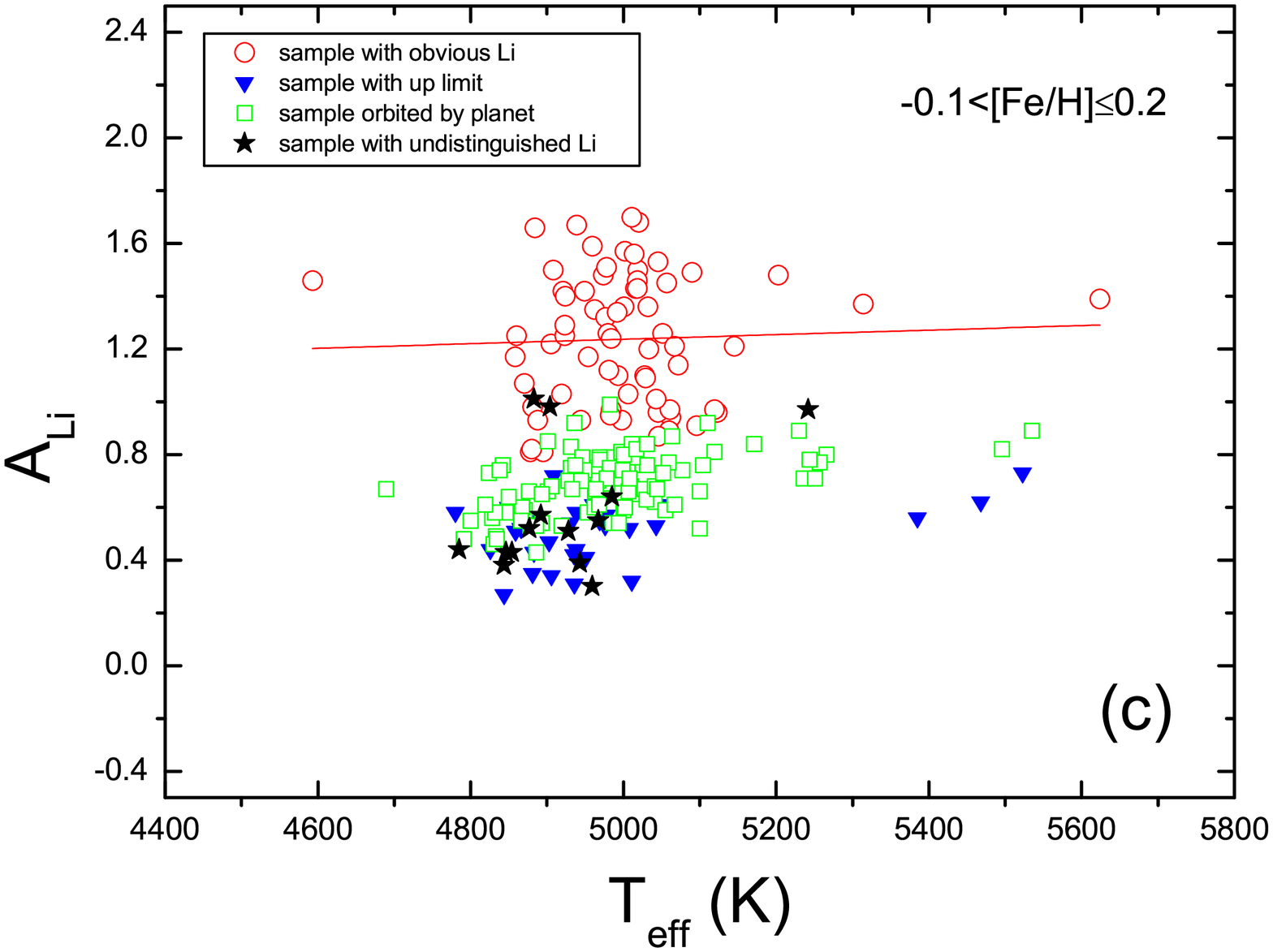}
   \caption{Lithium abundance against effective temperature in three different metallicity coverage.
   (a). -0.8 $<$ [Fe/H] $\leq$ -0.4, (b)-0.4 $<$ [Fe/H] $\leq$ -0.1, (c) -0.1 $<$ [Fe/H] $\leq$ 0.2.
   The symbols are the same as in Figure2.}\label{Fig. 9}
\end{figure}

\subsection{Lithium abundance versus rotational velocity}
The rotational velocities in this analysis for 321 giants are taken
from Takeda et al. (2008), but the rotational velocity are not
available for the other 57 giants in Liu et al. (2010). For these 57
giants, the total macrobroadening function is a convolution of
instrumental broadening, rotation and macroturbulence, and can be
derived from the process of deriving the lithium abundance, with a
maximum value of 8 km s$^{-1}$. Therefore, the rotational velocity
should be no faster than 8 km s$^{-1}$. Considering that there is no
systematic difference between a sample of 321 and 378, we only
compare the lithium abundance versus rotational velocity for the 321
sample giants in Figure 10. Since our sample is taken from the
planet search program, which select stars with slow rotational
velocity to ensure the spectral line will not be broadened too much,
the rotational velocity for most of our sample is slower than 6 km
s$^{-1}$, and there are only a few between 6-11 km s$^{-1}$.
Therefore, it is not surprising that the lithium abundance shows no
correlation with rotational velocity when it is smaller than 10 km
s$^{-1}$. The spread in lithium abundance is large in the values of
vsini $<$ 10 km s$^{-1}$, and only two stars show 10 km s$^{-1}$ $<$
vsini $<$ 11 km s$^{-1}$ in this sample. One of the latter presents
the highest lithium abundance and the other has only been assigned
with an upper limit. HD65228, the most Li-rich object, shows the
highest rotational velocity of 10.28 km s$^{-1}$, and it also shows
the highest temperature in our sample. Previous studies (De Medeiros
et al. 2000, de Laverny et al. 2003, L$\grave{e}$bre et al. 2006)
found that the stars with higher or moderate rotational velocity
(vsini $>$ 10 km s$^{-1}$) tend to show higher lithium abundances.
Our results are consistent with the results from de Laverny et al.
(2003), in which lithium abundance shows a wide spread at vsini $<$
4 km s$^{-1}$.

\begin{figure}
\epsscale{.70}
  \plotone{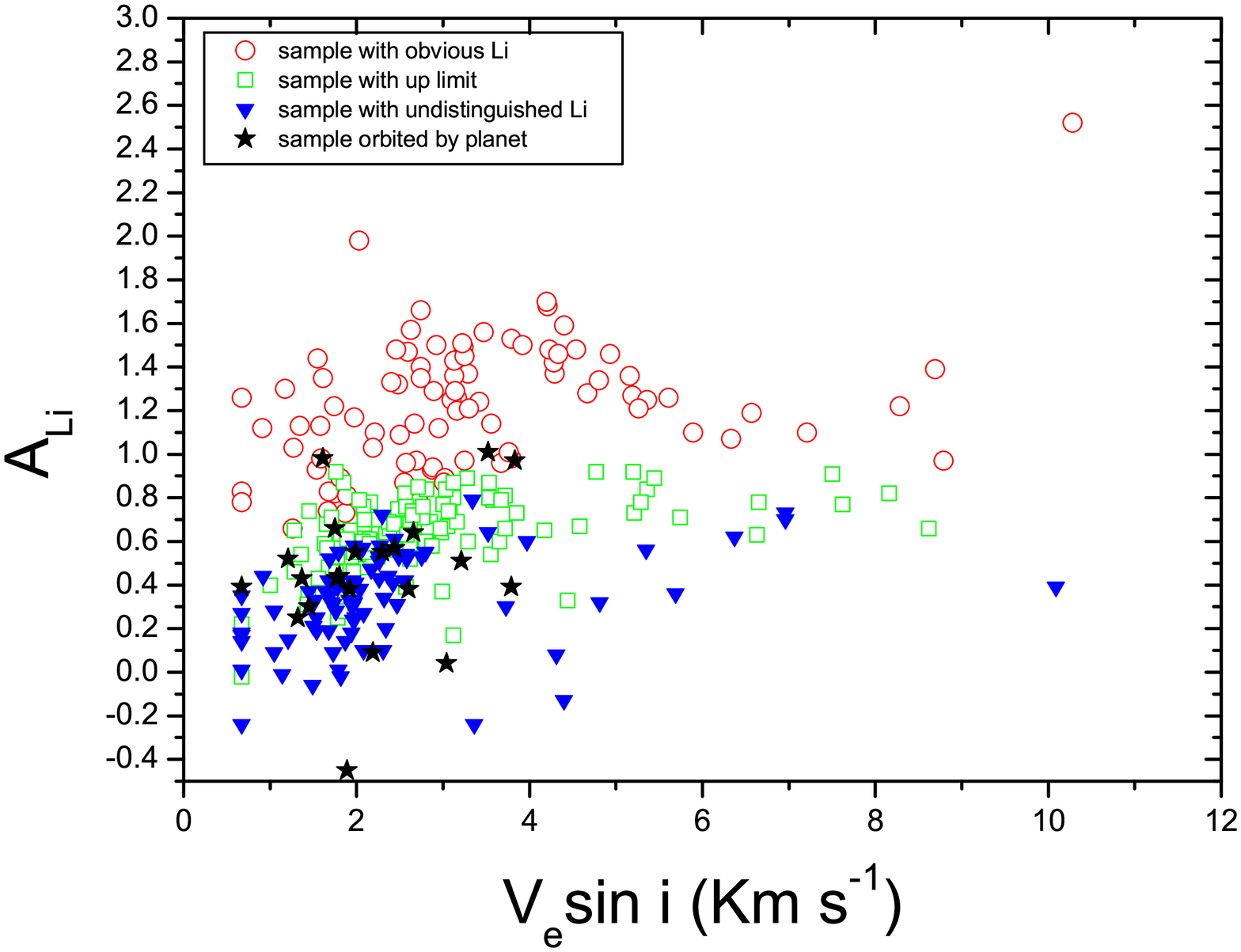}
   \caption{Lithium abundance versus rotational velocity. The symbols
   are the same as in Figure2.}
 \label{Fig.
10}
\end{figure}

Considering that many stars in our sample have an effective
temperature ranging from 4800-5100 K, we plot the
 lithium abundance for stars with the temperature of 4800-5100 K against metallicity and rotational velocity in Figure
 11, for the purpose of taking off the effect of effective
temperature. We omit the three lithium rich giants (A$_{\rm{Li}}$
$>$ 1.7) in Figure 11 to show the general behavior of lithium in red
giants. It becomes more clear that the lithium abundance is not
correlated with metallicity, or rotational velocity in the
temperature range of 4800-5100 K.
\begin{figure}
\epsscale{.50}
 \plotone{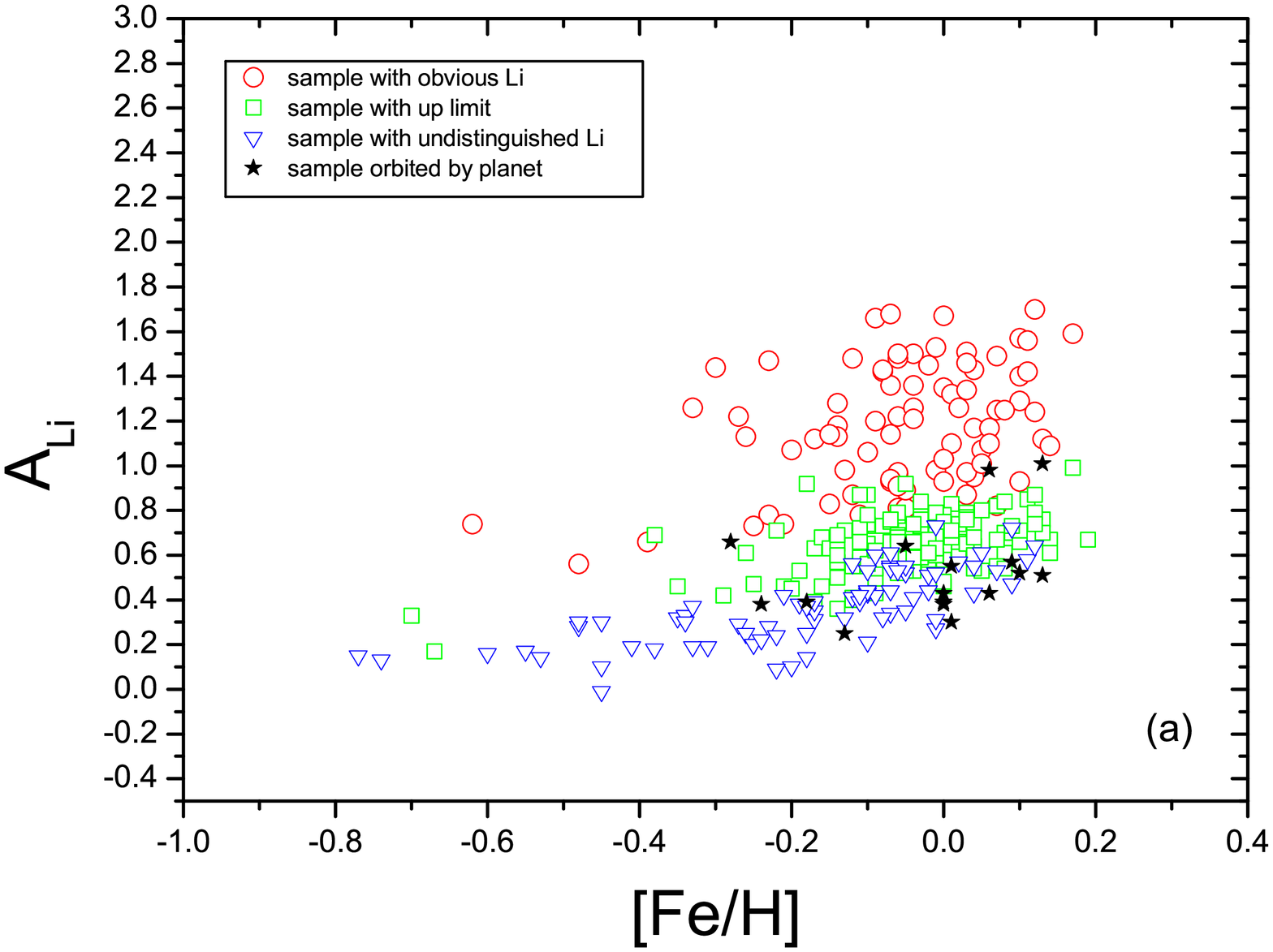}
 \plotone{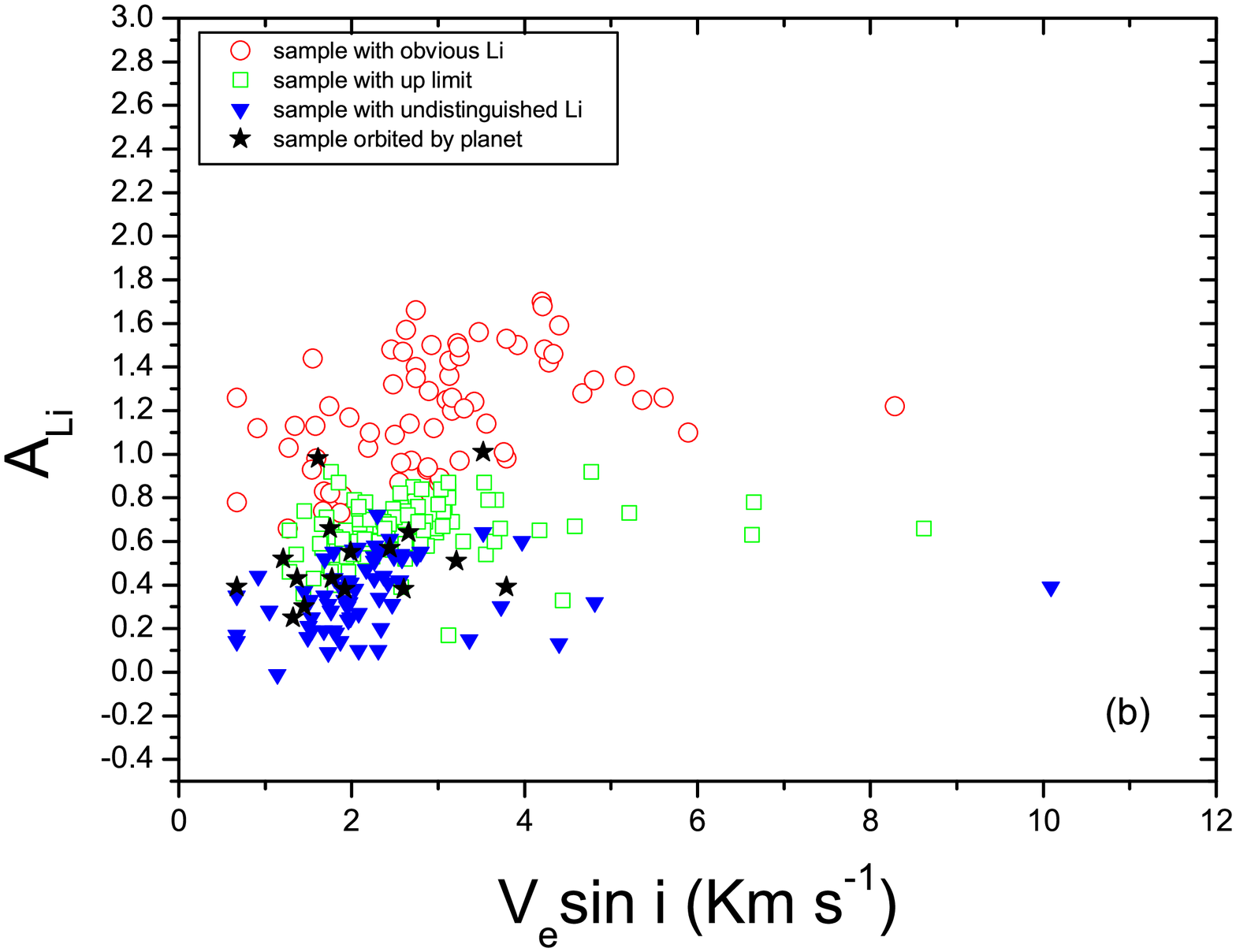}
   \caption{Lithium abundance against metallicity, and rotational velocity when 4800 $\leq$T$_{\rm{eff}}$ $\leq$ 5100 K.
  The symbols are the same as in Figure2.}\label{Fig. 11}
\end{figure}

\subsection{Lithium in planet-hosting giants and giants with not known giant planets}
Lithium abundances are believed to be important for potentially
providing a better understanding of processes involved in the
formation and evolution of planetary systems.

There are extensive studies on lithium in dwarfs that host planets
and those without giant planets; however, there are conflicting
results. It has been found in many studies (Israelian et al. 2004,
Takeda \& Kawanomoto 2005, Chen et al. 2006, Israelian et al. 2009)
that dwarfs that host planets, with effective temperature ranging
from 5600 to 5850 K, exhibit severely depleted Li. A large fraction
of the comparison sample only shows partially inhibited depletion.
However, some studies (e.g. Luck \& Heiter 2006, Baumann et al.
2010) have found that there is no difference in lithium abundance
between stars with and without planets, and they also claimed that
the previously found differences were attributed to different
stellar age and metallicity between stars that host planets and
comparison samples. Thus it is important to perform a comparison
between stars with and without planets with similar parameters
(metallicity and age), and in a consistent way.

Although many studies focus on whether lithium is more depleted, or
not, in dwarfs that host planets, such kind of investigations have
not been taken among giants. With more and more giants with planets
released by the Okayama and Xinglong planet search program, it is
now possible for us to explore the lithium behavior in giants with
and without planets in more detail. From Figure 5, which describes
the lithium behavior against effective temperature, a remarkable
feature appears: among 23 planet-hosting giants, only 8 stars have
Li being detected, with a maximin of A$_{\rm{Li}}$ = 1.01. However,
among the 355 stars which are not known to have planets, 271 stars
show Li detections. It can also be concluded from Figure 10(b-c)
that differences in metallicity from the stars with and without
giant planets cannot explain the diverse behavior of lithium
abundances. In addition, from the H-R diagram (Figure 7), the stars
with and without planets are located in the same regions, suggesting
that there is no difference in terms of stellar evolution for the
two sets of samples. The reason is probably due to the fact that
lithium is easy to deplete in stars with planets. This effect is
also unveiled among dwarfs with planets (e.g. Israelian et al. 2004,
Chen et al. 2006); however, this is the first time that this aspect
has been noticed for giants. Some studies(e.g. Siess \& Livio 1999,
Adam$\acute{o}$w et al. 2012) found that Li-enrichment may be caused
by planet accretion; however, there are no Li-rich giants in our 23
planet-hosting giants, which may suggest that such situation where
the accretion of a planet leads to the formation of a Li-rich giant
may be very rare. Since the number of giants with planets is limited
in this sample, more targets with planets are necessary for a more
detailed study.

All planet-hosting giants have vsini $<$ 4 km s$^{-1}$, which is due
to the fact that giants with lower rotational velocity are more
suitable for detection of planets with the radial velocity method,
since stars with rapid rotation normally show strong stellar
activity, thus there are severe variations in intrinsic RV.

\section{A large sample of lithium abundance}

To further check if the lithium abundance is a function of effective
temperature and metallicity, we supplement our results with the
lithium abundances from Luck07, based on the fact that the two
studies are consistent in the effective temperature and the lithium
abundances. There are 298 sample stars from Luck07, including 287
G/K giants and 12 A/F giants. Luck07 used high-quality data and the
spectrum synthesis method to determine the lithium abundances, which
adopted by this research is based on physical parameters. Except for
96 common stars, and 15 stars without physical parameters and their
corresponding lithium abundance, 187 stars are added to our sample
and the final sample contains 565 giants in total. Stars with clear
and undistinguished Li detections for the 378 giants are regarded as
stars with Li detections in this section.

Non-LTE correction is applied to the 187 added stars from Luck07
using the same method described in Section 3.3. As is shown in
Figure 4(b), there is no correlation between effective temperatures
in the two studies for the common stars, therefore, we adopt the
average value of the two works. For the metallicity in common stars,
there is a slight offset, with [Fe/H]$_{\rm{Liu2013}}$ = 0.04 $\pm$
0.01 + (0.97 $\pm$ 0.04) $\times$ [Fe/H]$_{\rm{Luck07}}$. There is a
slight offset of non-LTE lithium abundances for stars with Li
detections in both studies as well, and the correction function is
A$_{\rm{Li-Liu2013}}$ = 0.206 $\pm$ 0.03 + (0.845 $\pm$ 0.04)
$\times$ A$_{\rm{Li-Luck07}}$. Therefore, we correct the metallicity
and lithium abundance (only for stars with Li detections, and adopt
the original value for stars with upper limits) from Luck07 to ours
for the 187 added stars.

As Luck07 has not provided the UVW data, we calculate the kinematic
parameters (U$_{LSR}$, V$_{LSR}$, W$_{LSR}$) based on the method
given by Johnson $\&$ Soderblom (1987) for all these 567 stars with
radial velocities (which are already corrected to heliocentric
velocities) taken from Takeda et al. (2008), Liu et al. (2010) and
Luck07. The position, parallax, and proper motion are taken from
Hipparcos data. A Solar motion of (U, V, W)$_{\bigodot}$ = (-10.00
$\pm$ 0.36, +5.25 $\pm$ 0.62, +7.17 $\pm$ 0.38) in km s$^{-1}$
(Dehnen $\&$ Binney 1998) is adopted, where the Galactic velocity
component U is defined to be positive towards the Galactic
anticenter. The Galactic velocity components U, V, and W of stars in
our program are corrected to the local standard of rest (LSR). The
relative probabilities for the thick-disk-to-thin disk (TD/D)and
halo-to-thick (H/TD) membership are determined using the method
proposed by Bensby et al. (2003). To assign a given star to one of
these groups, following the suggestion of Bensby et al. (2005), we
adopt TD/D $\geq$ 2 for thick-disk stars, and TD/D $\leq$ 0.6 for
thin-disk stars, and all remaining stars with a probability between
these values are considered to be a member of the transition
population. Among the 565 stars in the enlarged catalog, there are
one star belongs to halo; 467 stars belong to the thin-disk; 43
belong to the thick-disk; and 54 are considered to be the transition
population. The only halo star HD39364, with H/TD = 1.52, that has
an upper limit in Li abundance will not be included in the following
comparison.

The lithium abundance as a function of metallicity and effective
temperature is shown in Figure 12, separately for thin- and
thick-disk stars. Black open dots, open triangles and stars
represent sub-samples with Li detections, upper limits, and planets
respectively for thin-disk stars, while red and blue dots, triangles
and stars represent those for thick-disk stars and the transition
population. In general, the lithium abundances increase towards
higher effective temperature and metallicity. The enlarged sample
includes 17 Li-rich giants, among which 14 are taken from Luck07,
containing 9 F giants, and 5 G/K giants. HD 194937 and HD 214995 are
K giants, which had been studied in Kumar et al. (2011), indicating
that the enrichment may be attributed to the conversion of $^{3}$He
via $^{7}$Be to $^{7}$Li by the Cameron-Fowler mechanism. As
discussed in Section 4.2, combining their positions in H-R diagram
(Figure 13), the three G giants are also supposed to be enriched by
the Cameron-Fowler mechanism. Excluding the Li-rich stars, the
lithium abundance is slightly increasing with increasing metallicity
and higher effective temperature for G/K giants. Different patterns
are shown for the two disks in this sample, indicating that the
lithium abundances in thick-disk stars are lower than those in
thin-disk in general. Meanwhile, excluding the one hot giant, there
are no Li-rich and very few Li-normal stars in the thick-disk
population. The different behaviors of thin-/thick-disk stars
indicate that the lithium abundances in thick-disk stars are more
depleted than those in thin-disk stars. The trends agree with the
theoretical prediction, which expects that the thick-disk giants are
older than thin-disk stars and their lithium abundances are easier
to become depleted due to the higher temperature inside the giants
along the RGB. The behavior of lithium abundances for giants in our
sample is in agreement with the results of dwarfs and subgiants
(Ram\'{\i}rez et al. 2012). No notable difference has been
identified between stars with and without planets for thick-disk and
thin-disk stars. From the metallicity distribution, the metallicity
for thin-disks star are lower than those in thick-disk, supporting
the theoretical prediction that thick disk stars are older. The
lithium abundance, UWV value and TD/D for 565 giants are listed in
Table 5, which is only available in electronic format.

\begin{figure}
\epsscale{.50} \plotone{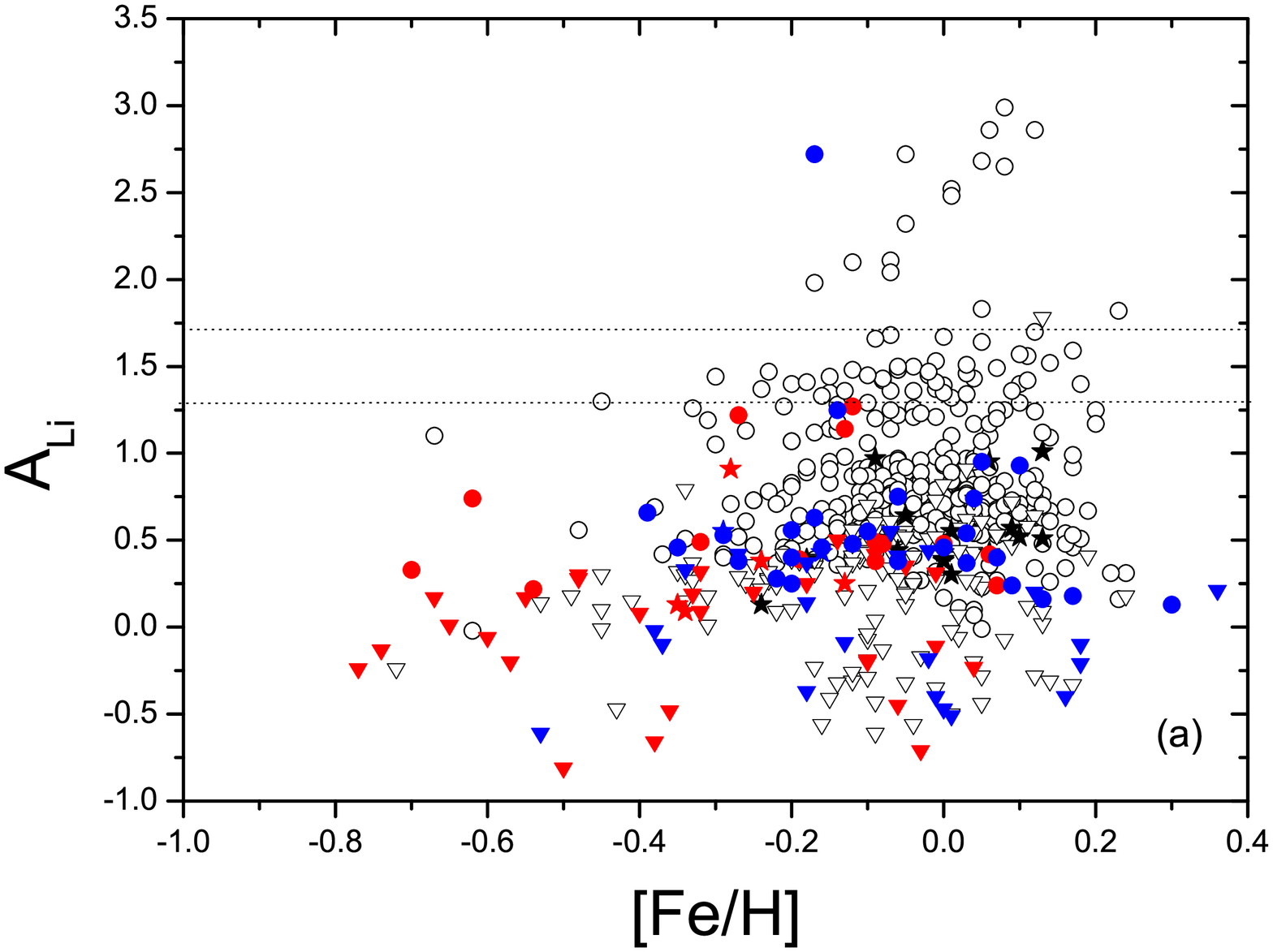} \plotone{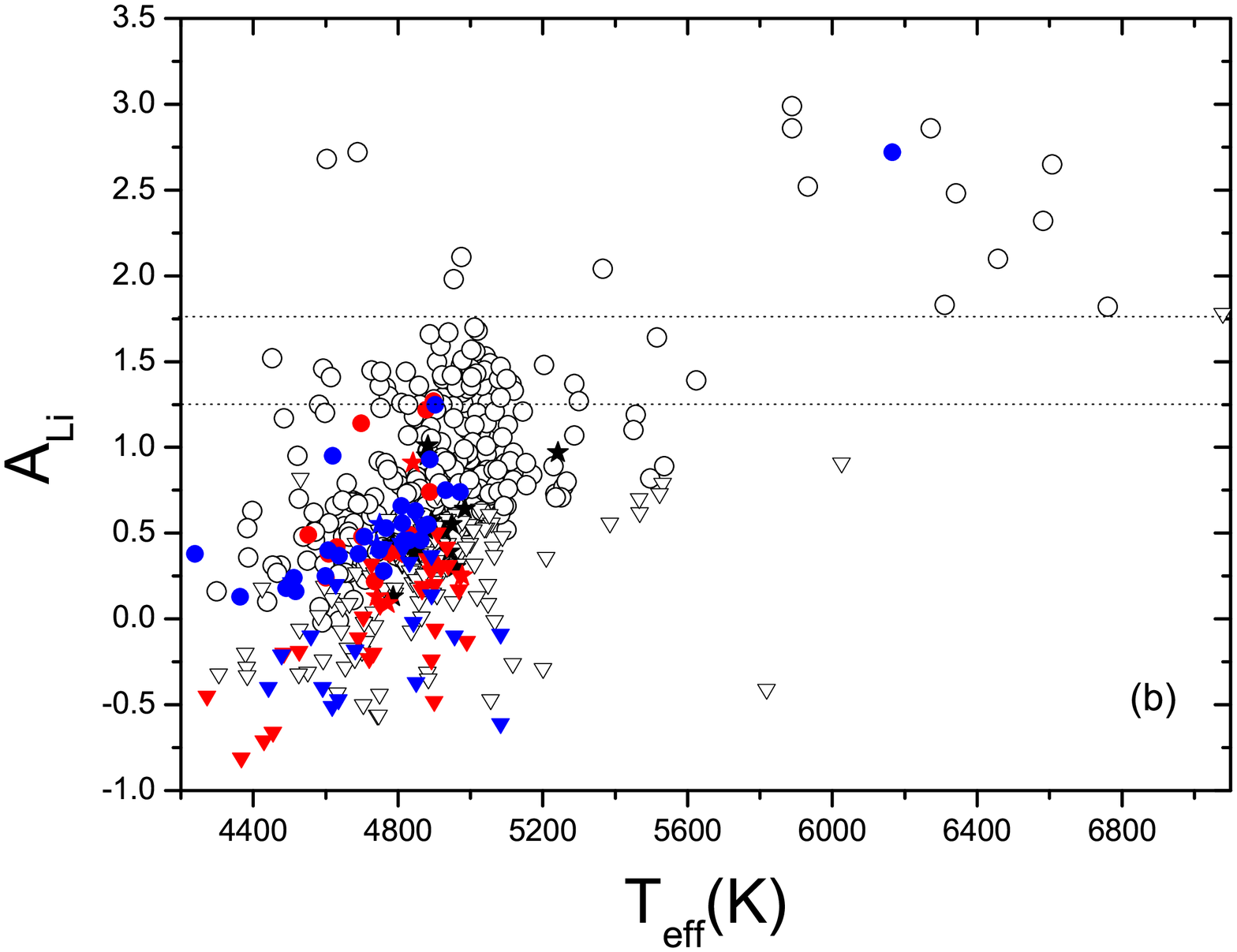}

   \caption{Lithium abundance against metallicity and effective temperature. The black open dots, open triangles and stars
represent sub-sample with Li-detections, upper limits and planets
for thin-disk, while red and blue dots, triangle and stars represent
those for thick-disk stars and transition population. \label{Fig.
12}}
\end{figure}

\begin{figure}
\epsscale{.50} \plotone{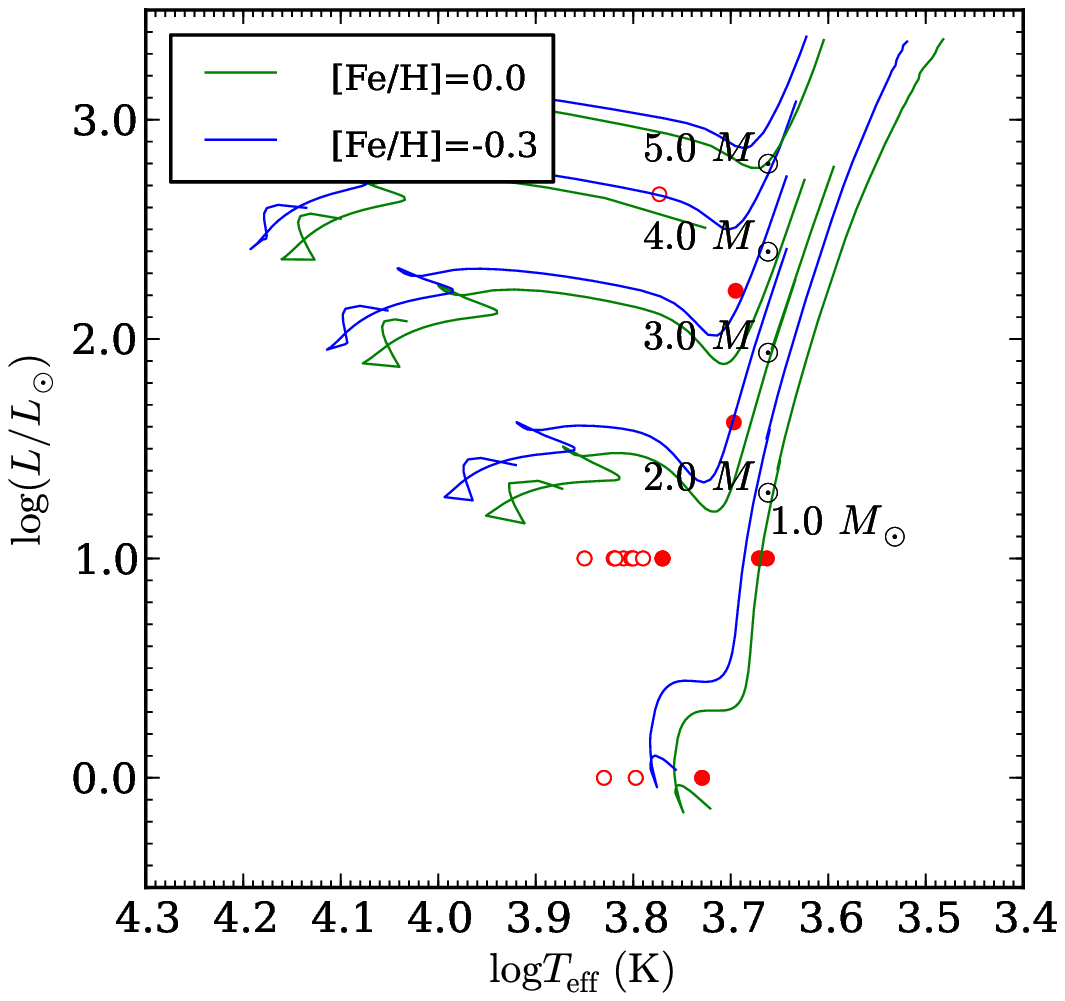}
   \caption{Lithium abundance for 17 Li-rich stars in H-R diagram. Open dots and filled dots represent giants with a spectral type of F and G/K.
\label{Fig. 13}}
\end{figure}

\section{Summary}
We present the lithium abundances for 378 intermediate-mass late G/K
giants by the spectrum synthesis method, among which 321 stars come
from the Okayama planet search program and 57 stars from the
Xinglong planet search program. The non-LTE correction is performed,
resulting in a correction of 0.05-0.28 dex, with the majority around
0.18 dex. The lithium abundance as a function of metallicity,
effective temperature and rotational velocity are
discussed. The main results are summarized as follows: \\

1. Our sample includes three Li-rich giants, 36 Li-normal stars and
339 Li-depleted stars. The fraction of Li-rich stars in this sample
agrees with previous works, which predict that less than 1$\%$
giants are Li-rich. Stars with normal amounts of Li are supposed to
present the abundance of current interstellar medium and have
experienced the expected dilution phase during the first dredge-up
phase. For the 339 Li-depleted stars, the abundance of Li is
depleted far more than the theoretical prediction, suggesting that
the Li deficiency may have already taken place in the main sequence
stage for many 1.5-5 M$_{\odot}$ stars which has been
 carried over to the evolved G/K giants.\\

2. The lithium abundance is a function of effective temperature and
metallicity, but for stars within the temperature range of 4800-5100
K, there is no clear correction between lithium abundance
and effective temperature.\\

3. The lithium abundance is not correlated with rotational
velocity when it is smaller than 10 km s$^{-1}$, which is consistent
with the results of previous studies. \\

4. For giants with planets, the lithium abundance is easy to
deplete. Stars with and without planets are located in the same
region on the H-R diagram, suggesting that there is no difference
\textbf{on} the stage of stellar evolution for stars with planets
and those without planets. All giants harboring planets have vsini
$<$ 4 km s$^{-1}$. The reason that planet-hosting giants show
smaller rotational velocity is due to the fact that lower rotational
velocity shows a suitable pattern to detect planets using the radial
velocity method.

In Section 5, we provide a catalog of stellar parameters, lithium
abundances for 565 giants, with supplementary of 187 stars from
Luck07. These data are used to investigate the lithium behavior as a
function of metallicity and effective temperature, and the
differences in lithium behavior for thin-/thick-disk stars. The
lithium abundances slightly increase with metallicity and effective
temperature in this enlarged sample of G/K giants, and thick-disk
stars present lower lithium abundance than thin-disk stars,
reflecting different degrees of lithium depletion in these two
populations.

 \acknowledgments

This research is based on data collected at Okayama Astrophysical
Observatory (OAO), which is operated by National Astronomical
Observatory of Japan (NAOJ). We are grateful to all of the staffs of
OAO for their support during the OAO observations. This work was
funded by the National Natural Science Foundation of China under
grants 11173031, 11233004, 11078022, 11103034, 11103030 and
11390371.

\end{document}